\documentclass[%
 aip,
 amsmath,amssymb,
 reprint,%
]{revtex4-1}

\usepackage{graphicx}
\usepackage{dcolumn}
\usepackage{bm}
\usepackage{amsmath}
\usepackage[utf8]{inputenc}
\usepackage[T1]{fontenc}
\usepackage{mathptmx}
\usepackage{etoolbox}
\usepackage{color}
\usepackage{booktabs}
\usepackage{multirow}
\usepackage{scalefnt}

\makeatletter
\def\@email#1#2{%
 \endgroup
 \patchcmd{\titleblock@produce}
  {\frontmatter@RRAPformat}
  {\frontmatter@RRAPformat{\produce@RRAP{*#1\href{mailto:#2}{#2}}}\frontmatter@RRAPformat}
  {}{}
}%
\makeatother
\begin{document}

%\preprint{AIP/123-QED}

\title{The starting dates of COVID-19 multiple waves}

\author{Paulo Roberto de Lima Gianfelice}
\affiliation{Federal University of São Paulo, São José dos Campos, 12247-014, Brazil}

\author{Ricardo Sovek Oyarzabal}%
\affiliation{Federal University of São Paulo, São José dos Campos, 12247-014, Brazil}

\author{Americo Cunha Jr}%
\affiliation{Rio de Janeiro State University, Rio de Janeiro, 20550-900, Brazil}

\author{Jose Mario Vicensi Grzybowski}%
\affiliation{Federal University of Fronteira Sul, Erechim, 99700-970, Brazil}

\author{Fernando da Conceição Batista}%
\affiliation{Polytechnic Institute of Leiria, Leiria, 2411-901, Portugal}

\author{Elbert E. N. Macau}%
\affiliation{Federal University of São Paulo, São José dos Campos, 12247-014, Brazil}
 \email{elbert.macau@unifesp.br}

\date{\today}

% --------------------------------------------------------------

%This infectious disease has proven to be very peculiar, not only because of the great speed of its spread, which generated a chain of transmission at a global level in a few months but also because of the high frequency with which highly infectious variants have emerged and spread, which may have multiple waves of contagion in the same geographic location as a side effect. 
%, so that it is not only possible but likely that previous infections have gone unnoticed since it is a large city with a very dense international air network and a large influx of tourists and travelers.
%Our results support that COVID-19 probably started to be disseminated locally in Rio de Janeiro as early as January 2020, before community transmission was officially registered and the appropriate control measures were carried out.

\begin{abstract}
% rev p/ Americo
The severe acute respiratory syndrome of coronavirus 2 (SARS-CoV-2) spread globally very quickly, causing great concern at the international level due to the severity of the associated respiratory disease, the so-called COVID-19. Taking Rio de Janeiro city (Brazil) as a studying case, the first diagnosis of this disease in occurred March 2020, but the exact moment when the local spread of the virus started is uncertain as the Brazilian epidemiological surveillance system was not widely prepared to detect suspected cases of COVID-19 at that time. Improvements in this surveillance system occurred over the pandemic, but due to the complex nature of the disease transmission process, specifying the exact moment of emergence of new community contagion outbreaks is a complicated task. This work aims to propose a general methodology to determine possible start dates for the multiple community outbreaks of COVID-19, using for this purpose a parametric statistical approach that combines surveillance data, nonlinear regression, and information criteria to obtain a statistical model capable of describing the multiple waves of contagion observed. The dynamics of COVID-19 in the city of Rio de Janeiro is taken as a case study, and the results suggest that the original strain of the virus was already circulating in Rio de Janeiro city as early as late February 2020, probably being massively disseminated in the population during the carnival festivities.
\end{abstract}

\maketitle
% --------------------------------------------------------------

% --------------------------------------------------------------
\begin{quotation}
The first infection by the SARS-CoV-2\cite{WU_F2020,Zhu2020p727,CoronaviridaeNature2020,Andersen:2020} coronavirus was reported in Wuhan, Hubei Province, China, in late December 2019, with the spread of the {COVID-19} disease~\cite{WHO2020-02-11,The_Lancet} across China's borders happening very quickly~\cite{WHO2020-01-21}. Within a few months, COVID-19 has spread to many countries, which led the World Health Organization (WHO) to declare a pandemic state on March 12th, 2020\cite{WHO2020-03-11}. This pandemic has been recognized as one of the most serious public health crises faced by humanity in recent history, on an interval fewer than two years, deaths officially registered around the World already add up to more than 5.65 million lives\cite{JHU-COVID-19,WORLD-METERS} until this submission, a health catastrophe not seen since the Spanish flu pandemic \cite{Taubenberger2006} in the early 20th century.  A key question in this context is to find out, based on reported surveillance data, the moment in time when this disease start inside a community in other parts of the World. The answer to this question may lead to a better understanding of the epidemic propagation routes, and so, can help to establish strategies to refrain future outbreaks.
\end{quotation}
% --------------------------------------------------------------

% --------------------------------------------------------------
% rev p/ Americo
\section{Introduction}
\label{sec:intro}

\begin{figure}
    \centering
    \includegraphics[scale=0.2]{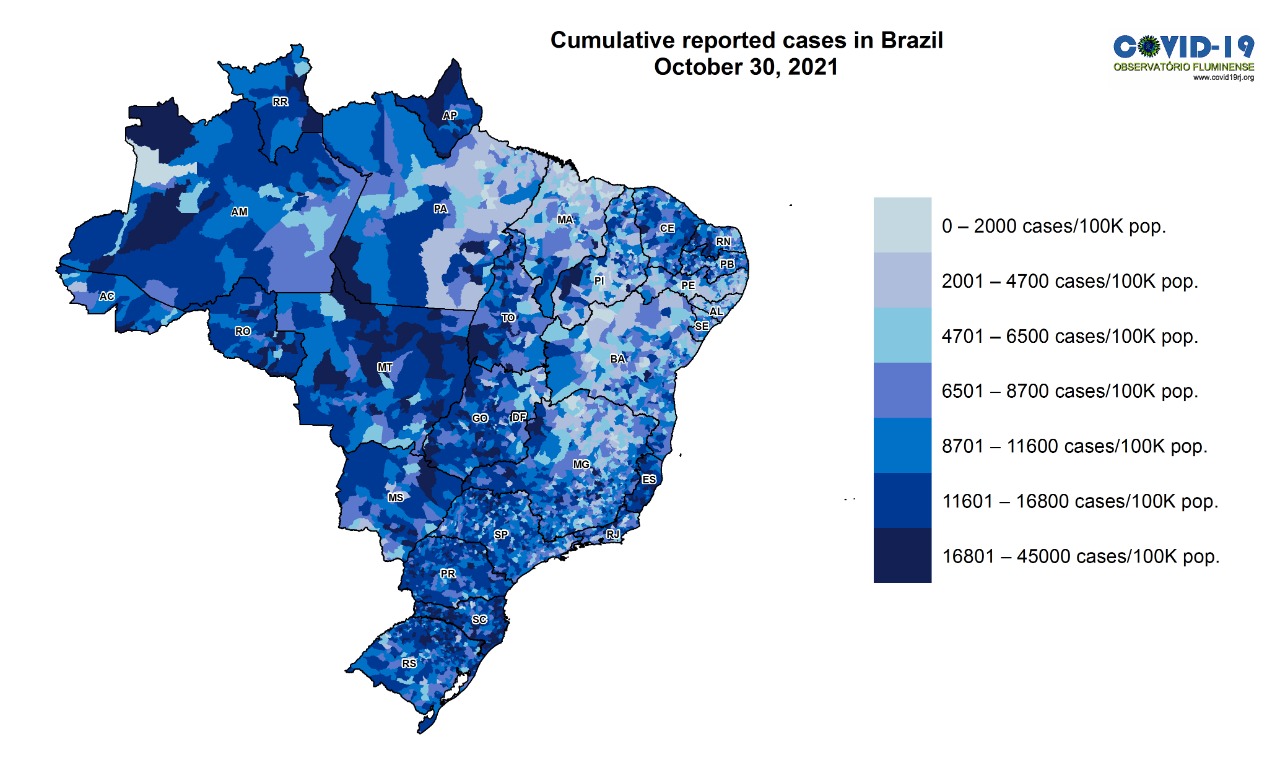}
    \vspace{2mm}
    \includegraphics[scale=0.2]{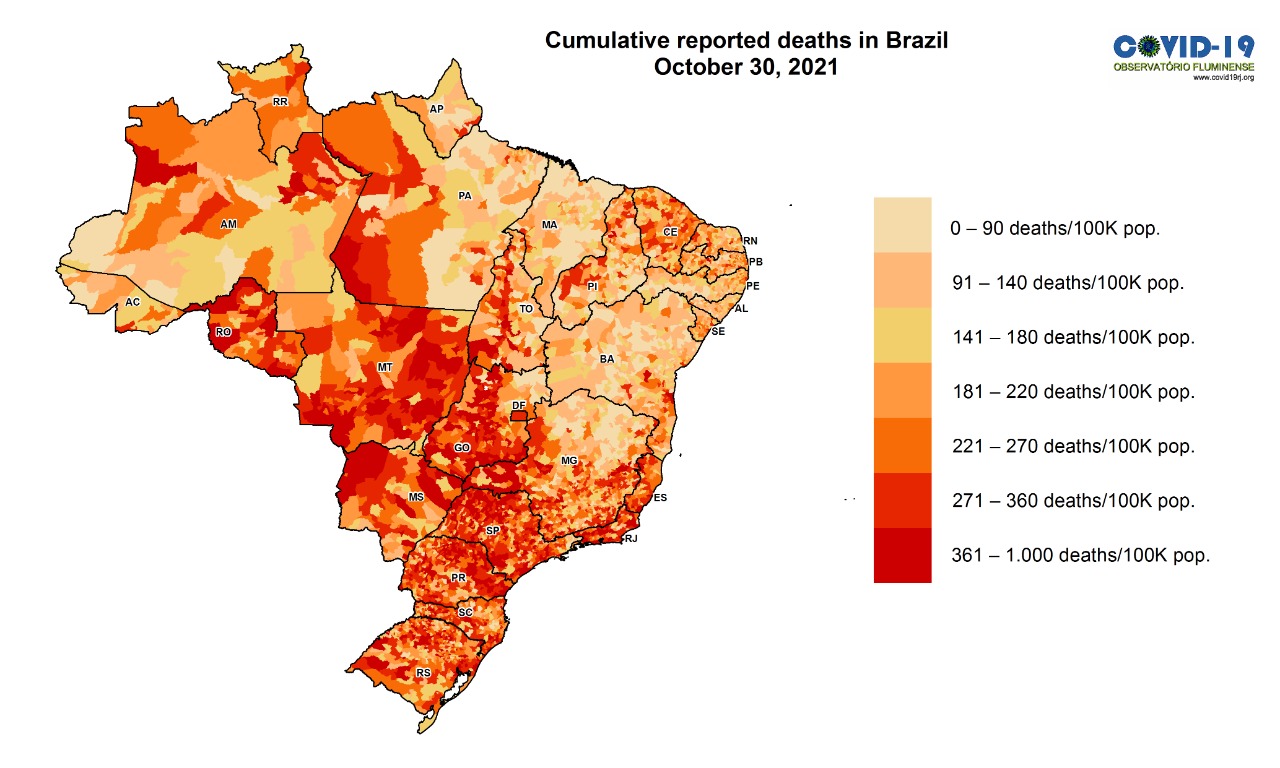}
    \caption{Prevalence of COVID-19 cases and deaths confirmed in Brazil between March 2020 and October 2021 \cite{OF-COVID19-Relatorio30}.}
    \label{fig:fig1}
\end{figure}

\begin{figure}
    \centering
    \includegraphics[scale=0.2]{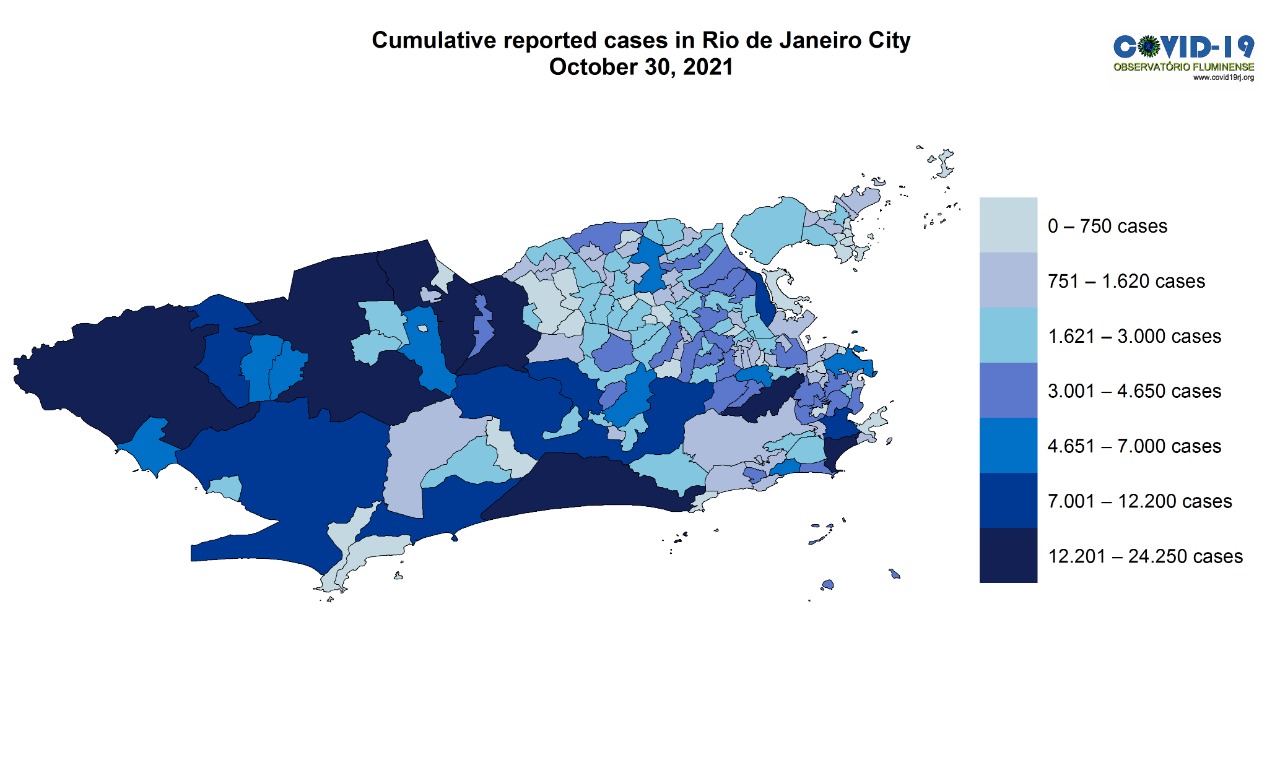}
    \vspace{2mm}
    \includegraphics[scale=0.2]{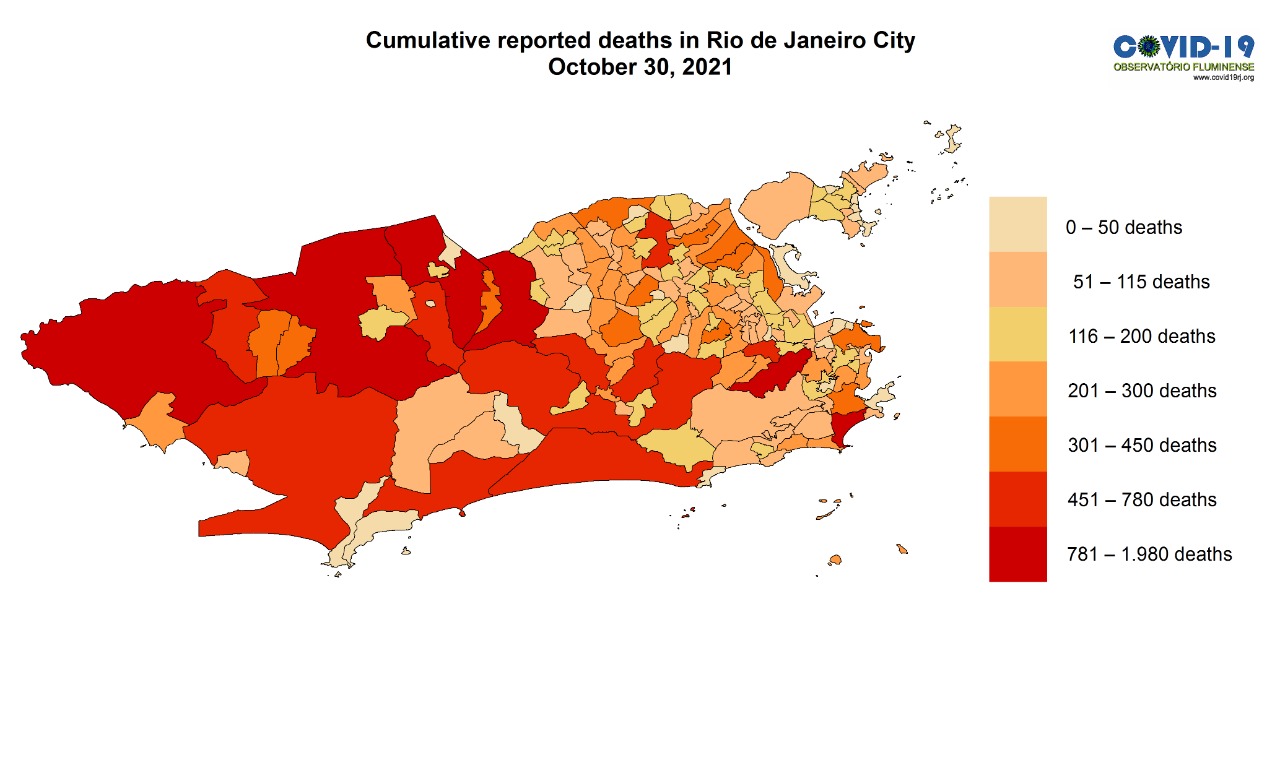}
    \caption{Prevalence of COVID-19 cases and deaths confirmed in Rio de Janeiro city between March 2020 and October 2021 \cite{OF-COVID19-Relatorio30}.}
    \label{fig:fig2}
\end{figure}

Brazil reported COVID-19 first official case on February 25, 2020\cite{AgenciaBrasil2020_1}, in the city of São Paulo, which is the largest municipality in the country. Similar notifications in other Brazilian cities quickly followed, being accompanied by a devastating scenario of contagion and death \cite{Lovisolo2020,OF-COVID19-Relatorio30}, as shown in Fig.~\ref{fig:fig1}, which presents the prevalence of cases and deaths in the country until October 2021. The first official notification in Rio de Janeiro, the second largest and most visited city by tourists, occurred on March 6, 2020\cite{Cavalcante2020}, with the epidemiological situation quickly evolving to one of the worst scenarios at the national level, among the highest number of confirmed cases and deaths by COVID-19, as can be seen in the prevalence maps of Fig.~\ref{fig:fig2}.

\begin{figure*}
    \centering
    \includegraphics[scale=0.45]{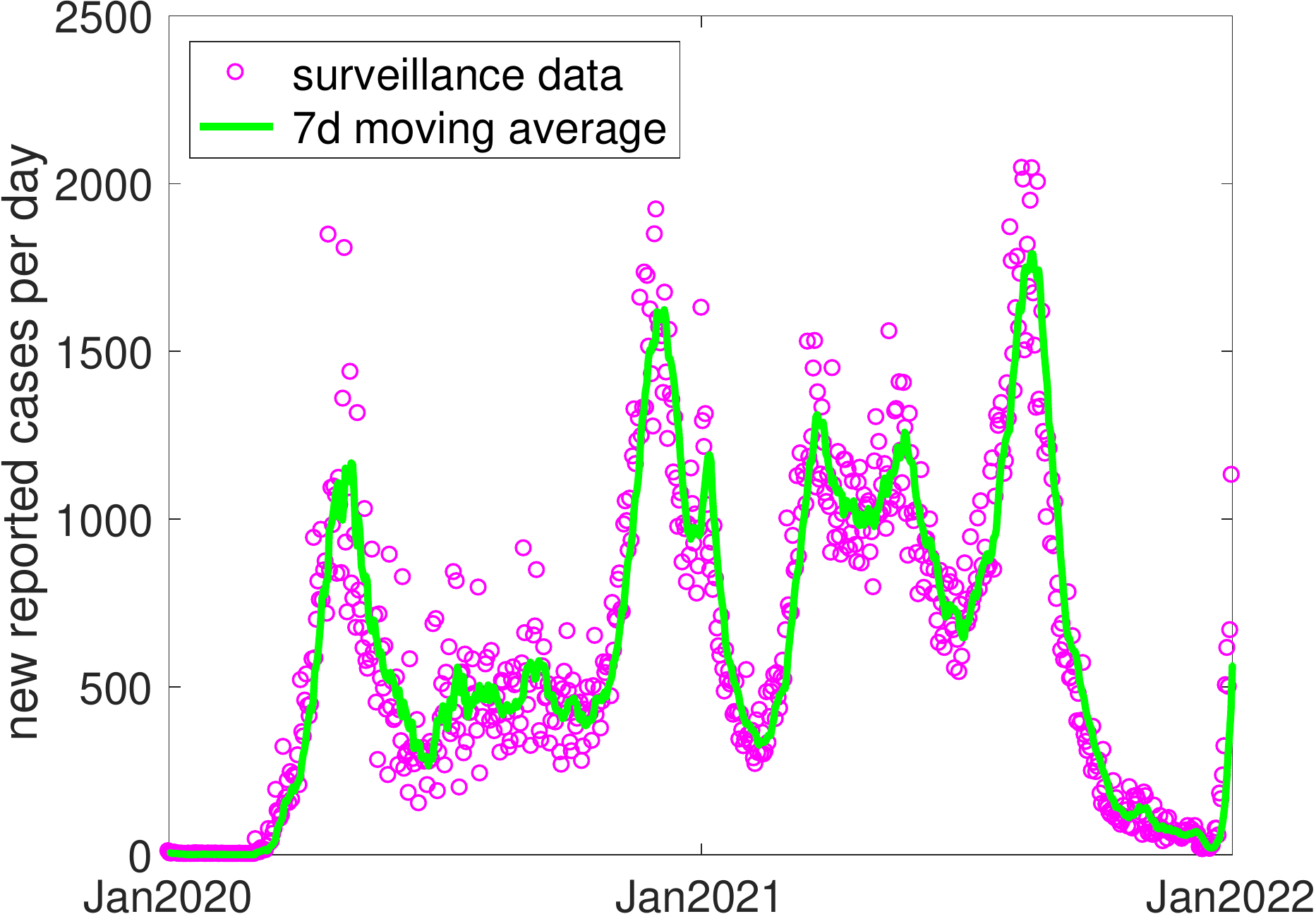}
    \includegraphics[scale=0.45]{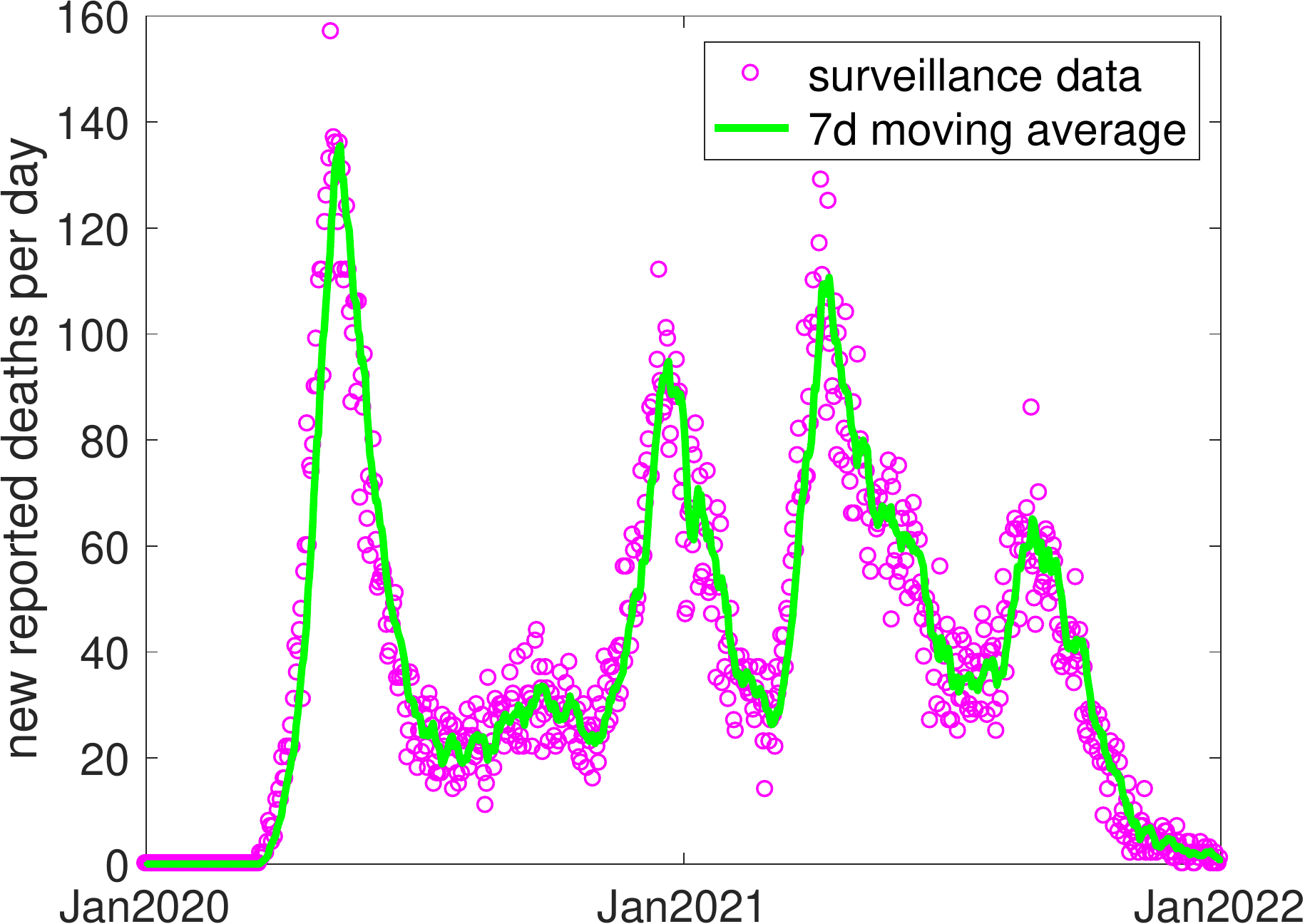}\\
    \vspace{5mm}
    \includegraphics[scale=0.45]{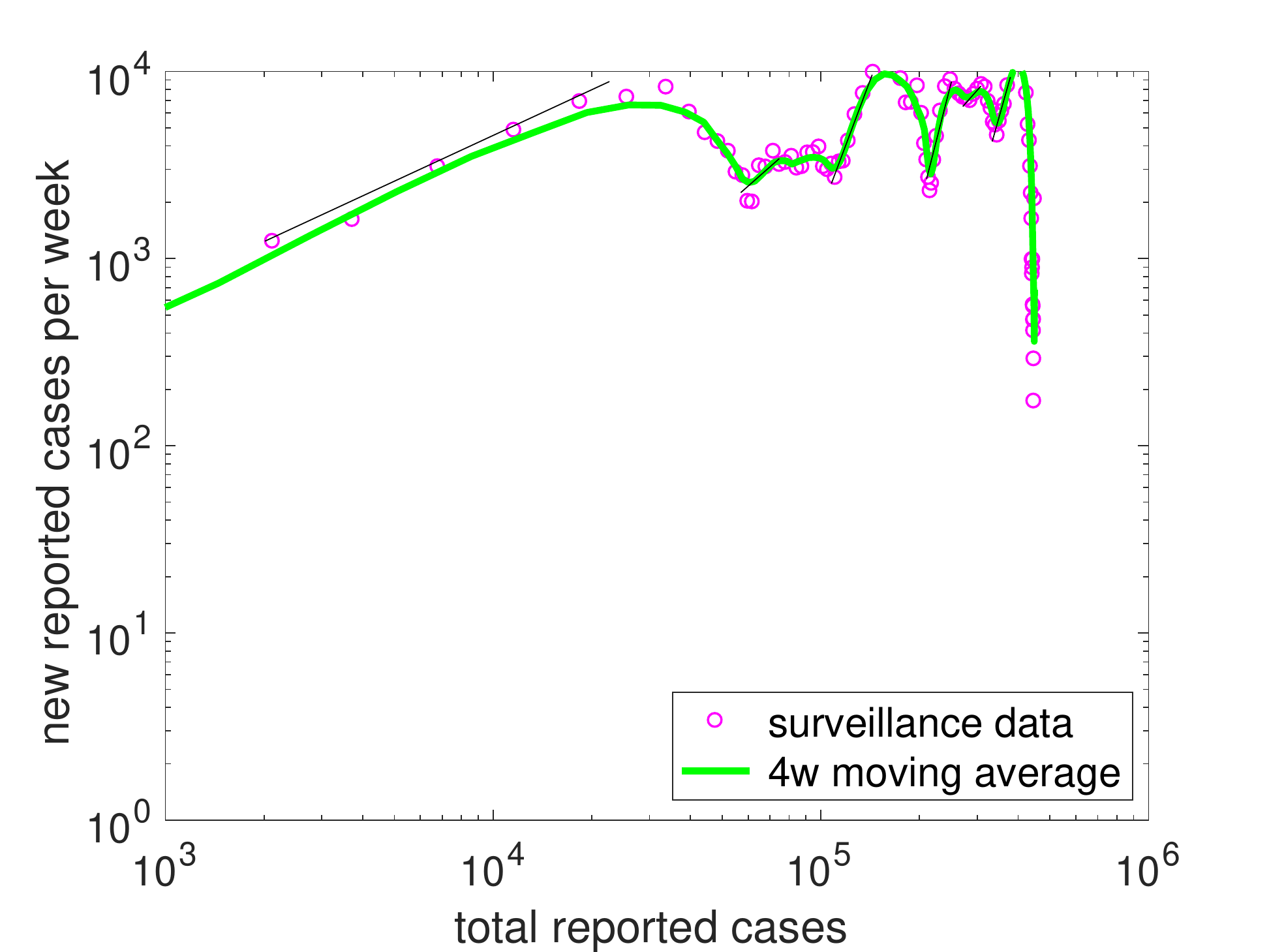}~~~~~~~~~
    \includegraphics[scale=0.45]{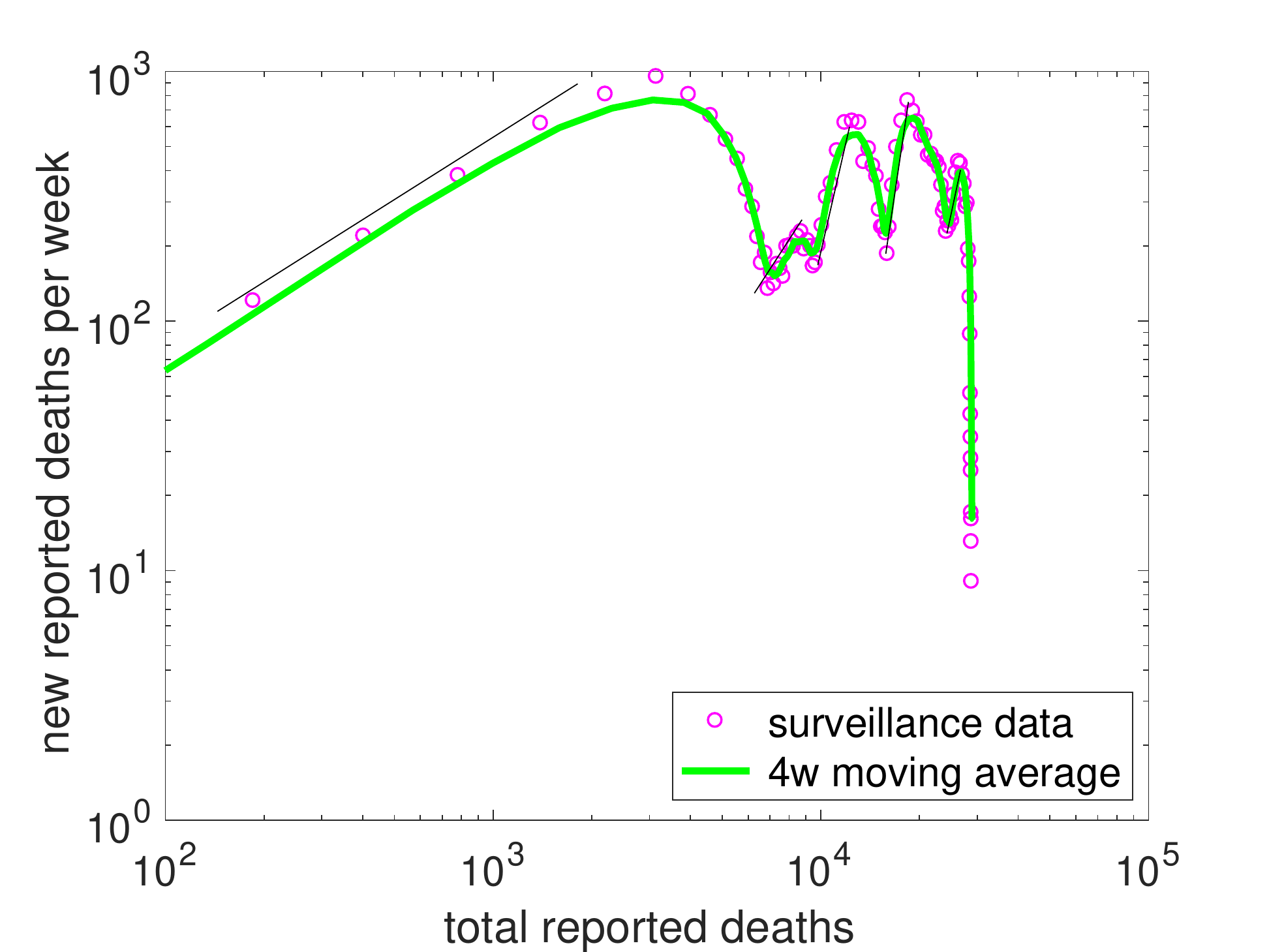}~~~~~
        \caption{Evolution of the incidence of confirmed cases (top left) and deaths (top right) of COVID-19 in Rio de Janeiro, from January 2020 to December 2021, organized by first symptoms date and event date, respectively. Also available is the comparison between new notifications per day and total notifications for cases (bottom left) and deaths (bottom right). The raw data are represented in magenta dots, and the 7 days moving average by the green curve. It can be noted the presence of multiple waves of contagion and deaths, where the four most pronounced peaks were in April 2020, December 2020, April 2021, and August 2021, respectively.}
    \label{fig:fig3}
\end{figure*}

Furthermore, Fig.~\ref{fig:fig3} shows the incidence of confirmed cases (top left) and deaths (top right) in Rio de Janeiro from January 2020 to December 2021, organized by first symptoms date and event date, respectively. It also shows a comparison between new notifications per week and total notifications for cases (bottom left) and deaths (bottom right), where periods of incidence proportional to prevalence (exponential growth) are noticed. In this figure, the raw data are represented in magenta dots and the 7 days moving average by the green curve. Multiple waves of contagion/deaths can be seen, the four most pronounced in April 2020, December 2020, April 2021, and August 2021, respectively. Small oscillations over time, which resembles an endemic period, can also be seen. In addition, the surveillance data for new confirmed cases (top left of Fig.~\ref{fig:fig3}) also highlights the existence of case notifications in January and February 2020, well before March 1st, the current date where it is already confirmed (through surveillance data) that community transmission had started. These cases, registered by the date reported for the first symptoms, were not counted at the beginning of March 2020 because their registration in the system only occurred a posteriori. For this reason, March 6th was announced that time as the day of the first confirmed case.  Although these early records may be due to errors in the registration, due to the provision of false or misleading information by patients, the fact that they are not isolated notifications may suggest that community transmission of SARS-CoV-2 in Rio de Janeiro city may have started well before the period close to March 1st, where community transmission is known to have already occurred.

As at the beginning of the pandemic the Brazilian epidemiological surveillance system was not prepared to massively register COVID-19 cases, it is not only possible but probable that the introduction of the virus in the community took place in a period earlier than of the current estimates for the starting date of community transmission (March 1st). The early cases mentioned above provide direct, albeit weak (due to possible recording errors or biases), evidence in this sense.

The SARS-CoV-2 fluctuation in Rio de Janeiro, as well as the significant number of notifications before the first official confirmation, are peculiar epidemiological phenomena, which differ from the typical situation observed in the majority of Brazilian cities. Thus, this situation deserves to be further investigated, since understanding this multiple wave behavior can provide insights into different stages of the epidemic spread, information that can be useful to guide decision-makers in future outbreaks.

In this context, one piece of information that is particularly interesting is the likely start date of each community outbreak, i.e.,
when the transmission inside a community is due to the locals. These dates are important because they can indicate key events, such as the introduction period of a new viral strain in the population\cite{Voloch2021, resendeetal}, or a drastic change in social behavior that is capable of inciting a new phase of epidemic expansion in a given population.

Several mathematical approaches can be used to access the dynamics of an epidemic wave, including the estimation of their initial date. Compartmental models\cite{Brauer2008,Brauer2017,Martcheva2015} based on differential equations are very natural for this purpose, with the underling parametrization obtained with aid of data-assimilation techniques, such as Kalman filter\cite{Rajnesh2021p1}, nonlinear regression\cite{Kucharski2020,Magal2020p3040,Subhas2020p071101,Lobato2021nody}, Bayesian statistics\cite{Cotta2020p220,Lyra2020}, neural networks and other machine learning tools\cite{Magri2020,He2020,ALBANI2021p6088,Kuhl2021}, etc. Such compartmental models are also fundamental in approaches that employ the concept of complex networks to describe the epidemic dynamics in a large population with heterogeneous spatial distribution\cite{Chen2020,Costa2020p043306,Aleta2020p964,Aleta2020p157,Ventura2021}. It is also quite common to describe the evolution of epidemic waves with aid of purely phenomenological models, supported by frequentist\cite{Liu2020p1527,Vasconcelos2020,Pelinovsky2020p110241,Vasconcelos2021,Vasconcelos2021SR,Chol-jun2022,Pelinovsky2022p111699,Contoyiannis2022p043109,Contoyiannis2022p011103} or Bayesian\cite{Romadhon2021,Calatayud2022} statistical approaches.

%\cite{Brockmann2013p1337,Maier2020p742}

There are already some works in the open literature concerned with estimating the start date of a local epidemic outbreak of COVID-19, such as Delatorre et al.\cite{DELATORRE}, which tries to infer the starting date of the SARS-CoV-2 spread in Western Europe and the Americas; Batista and Cunha~Jr\cite{Fernando_Batista}, that did the same exercise for the early stages of COVID-19 in Portugal; and Zhai et al.\cite{Zhai2021p013155}, which investigated the initial COVID-19 dynamics of 10 states in the U.S.A., New York City, United Kingdom, Italy, and Spain.

Despite the consistency of the results presented by the first two studies, by comparison with evidence (other epidemic observations) that support the estimated dates, they are more focused on reporting dates, having relatively few details about the underlying mathematical methodologies. The third one, on the other side, is rich in methodological detail, presenting a mechanistic framework based on a non-Markovian delayed compartmental model, which provides a relatively general tool for inferring the start date of an outbreak. However, in terms of validating the results, the authors' arguments implicitly rely heavily on the assumptions that the available data are sufficiently accurate and informative to calibrate the dynamic model, and that the latter provides a sufficiently realistic representation of the epidemic outbreak of interest. The scenario where at least one of these assumptions is not valid is not rare, so there is space in the literature for contributions that attack the same problem on other fronts.

Seeking to contribute to expanding the arsenal to estimate the initial date of an epidemic community outbreak, this paper discusses a generic parametric statistical approach for estimating the start date of an epidemic outbreak based on epidemiological surveillance data. The fundamental idea is to present a procedure that incorporates elements of generality, to be applied in typical epidemic outbreaks, but that is simple enough to be used by researchers who do not have strong training in epidemiology or more advanced statistical methods (e.g. Bayesian inference). In this way, we combine the surveillance data with algebraic multiple waves models, nonlinear regression, and information criteria to obtain a simple but representative mathematical model of the underlying outbreak, which provides an interval estimate for the start date of the referred contagion wave. The proposed methodology is illustrated with aid of COVID-19 data from Rio de Janeiro city, which has complex dynamics, with multiple contagion waves and (the typical) very irregular data.

The paper is organized as follows. Section~\ref{sec:methodology} describes the statistical framework. Section~\ref{sec:Results and discussion} presents the results of a case study in Rio de Janeiro. Finally, conclusions are shown in Section~\ref{sec:Conclusions}.
% --------------------------------------------------------------

% --------------------------------------------------------------
\section{Method}
\label{sec:methodology}

\begin{figure*}
    \centering
    \includegraphics[scale = 0.33]{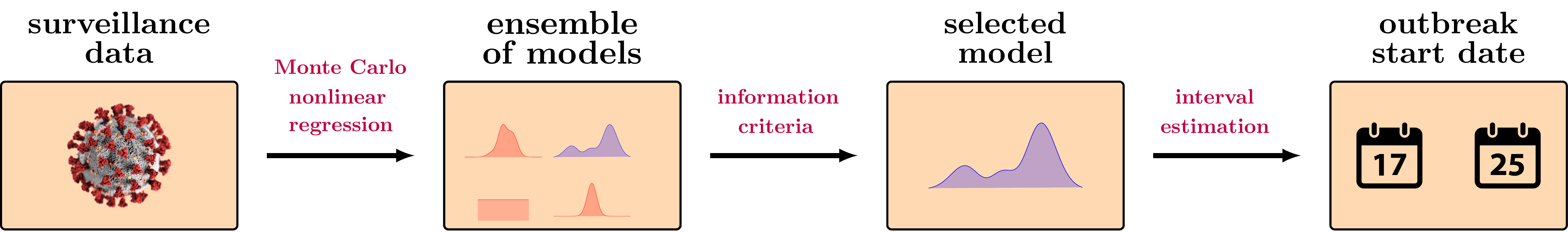}
    \caption{Statistical framework adopted to describe the COVID-19 outbreaks and infer the start period of each contagion wave. From epidemiological data, several statistical models are built with the aid of a linear regression process via Monte Carlo simulation. Information criteria are used to choose the most representative model(s), which is (are) used to provide an interval estimate for the start date of the COVID-19 contagion wave.}
    \label{fig:fig4}
\end{figure*}

The methodology employed here combines epidemiological surveillance data, algebraic statistical models, nonlinear regression, and a model selection procedure to infer the systematic behavior of COVID-19 outbreaks, focusing on estimating a possible start date for each contagion wave. A schematic version of this framework can be seen in Fig.~\ref{fig:fig4}, and each of its steps is described below.

\subsection{Surveillance data}

To monitor the evolution of COVID-19 spread, time series that quantify the new reported cases and deaths may be used. Very often such time series are obtained by a process of selection and agglutination of raw data from a large spreadsheet, where each line corresponds to a patient (duly anonymized), containing many dozens of information, such as zip code, date of first symptoms, and date of outcome (recovery or death), etc. These data are registered by the municipality health authorities. In the case of Rio de Janeiro city, they come from two databases: (i) e-SUS VE; and (ii) SIVEP-Gripe. The first records the flu syndrome in ordinary cases, while the second is responsible for the severe acute respiratory syndromes records. Both are compiled (without duplication) by the Municipal Health Department, which makes them available on a website\cite{PainelRioCovid} widely accessible to the general public. Data for new cases are organized by the date of first symptoms, while new deaths are recorded by the date of the event. Other locations have their registration systems, but the vast majority of them have most of the features in common with the database described above.

Epidemiological data are notoriously problematic for statistical analysis purposes, as delay and underreporting effects are unavoidable, due to the physical impossibility of knowing all the cases that occur, neither in real-time nor with a few days delay\cite{Gamerman2022}. In practice, as the city of Rio de Janeiro has not adopted a massive random testing policy, only part of the cases of infection by COVID-19 were confirmed with the laboratory tests, essentially those that seek medical attention. Thus, the available data for cases evolution provide a distorted (biased) picture of the epidemiological reality, with far fewer cases reported than the actual number of infected, and with an estimated date for infection beginning shifted in time from the true onset, since the infection onset is estimated by the date of first symptoms (which occurs on average 5 days after the infection event). Death records also suffer from underreporting, but on a much smaller scale, as in this case registration is compulsory. Typically, deaths from COVID-19 that are not counted are those with no formal diagnosis of the disease by a laboratory test. In this setting, the time series for the number of new deaths per day can be considered a proxy to assess the evolution of contagion by the disease, since there is a clear correlation between it and the temporal evolution of the total number of cases (part of the cases evolves to death, more cases occurring, the expectation is more deaths, fewer cases, fewer deaths)\cite{Gamerman2022}. 

In the analysis reported in this paper, only death data are considered. Since these data are recorded by the actual date of the event (in this case, the death), no delay is inserted into data, except that one related to the insertion of data into the system, which can take up to 30 days, but which only compromises the accuracy of the end of the time series, and can be corrected with the use of nowcasting techniques\cite{Gamerman2022}.

As the time series of cases and deaths are subject to much fluctuation, due to the natural variability of the disease, but also due to imperfections in the surveillance system, treating the data in advance is necessary to make them suitable for visual analysis, since extracting patterns from raw data is quite difficult. In this sense, these data are smoothed with the aid of 7 days moving average filter, to reduce excess fluctuation. As can be seen in Fig.~\ref{fig:fig3}, this treatment helps to display the data evolution pattern. But it should be mentioned that due to the use of only past data in the smoothing (for reasons of causality), the moving average inserts a delay of a few days in the data series. To avoid introducing this bias into statistical estimates, only raw data are considered in the statistical estimation processes presented below. Moving averages are used in this paper only to visually illustrate the trend of raw data.

\subsection{Ensemble of statistical models}

The three typical phases of an epidemic outbreak (expansion, transition, and exhaustion)\cite{brauer2001,Murray2002,Kuhl2021} usually are well described by logistic curves when the underlying population is relatively homogeneous, so that contamination comes out of a close interaction between two people. This scenario is assumed as a working hypothesis in this paper, in a way that several statistical models, based on the algebraic solution of a logistic differential equation, are used here to represent the multiple outbreaks of COVID-19 that are characterized by a dataset that collects deaths records for a certain locality into a time series. This is inspired by the approach adopted by Batista and Cunha~Jr\cite{Fernando_Batista} to study the early stages of COVID-19 in Portugal.

%Epidemic data evolution (both cases and deaths) usually is well described by logistic curves, as these can reproduce the three typical phases of an epidemic outbreak (expansion, transition, and exhaustion)\cite{brauer2001,Murray2002,Kuhl2021}. Thus, here we assume that we have a condition that the contamination among people is properly describer in the context of a homogeneous scenario in a population, so that the contamination comes out of a close interaction between two people.

 It means the we have the statistical

\subsubsection*{Single wave model}

When just a single wave of contagion is of interest, the Verhulst logistic model \cite{Martcheva2015,Brauer2008,Murray2002,brauer2001} is employed. This model assumes a growth rate proportional to the disease prevalence in the population at time $t$, denoted by $C(t)$, but with a constant of proportionality that changes in time, so that it reduces when $C(t)$ grows, until reaching a maximum sustainable population in the limit when $t \to \infty$. This logistic model is defined by the differential equation
\begin{equation}\label{log02}
                \dfrac{dC}{dt} =  r \, C \, \left(1 - \dfrac{C}{K}\right) \, ,
\end{equation}
which has as solution the classical logistic curve
\begin{equation}\label{Ct}
                C(t) = \displaystyle \frac{K}{1+e^{\displaystyle -r \, (t-\tau)}} \, ,
\end{equation}
where $r$ represents the infection growth rate, $K$ is the final number of notifications at the end of the outbreak, and $\tau$ describes the (initial condition dependent) instant of inflection associated to this curve, when the exponential growth ends, and a sudden deceleration begins. The derivative of $C(t)$, dubbed the incidence curve $I(t)$, represents the number of new death notifications per day, and is given by
 \begin{equation}
    I(t) = \dfrac{dC}{dt} = \frac{r \,  K \, e^{\displaystyle -r\, (t-\tau)}}{\left(1 + e^{\displaystyle -r\,(t-\tau)} \right)^2} \, .
\label{It}
\end{equation}

Other authors use generalized logistic curves to deal with this kind of epidemic data \cite{Vasconcelos2020,Fernando_Batista,Zou2020p1}. Despite this being a possibility, the present work opted for the classic logistic curve for the sake of parsimony, as it is capable of providing a reasonable representation of Rio de Janeiro outbreaks.

\subsubsection*{Multiple waves model}

When more than one wave of contagion matters, a multimodal epidemic curve with $N$ peaks is considered, which demands the estimation of $3N$ parameters
\begin{equation}
\theta = (K_1,\, r_1,\, \tau_1,\, \cdots , \, K_N,\, r_N,\, \tau_N) \, ,
\label{theta_eq}
\end{equation}
resulting in a global prevalence curve
\begin{equation}
    C(t) = \displaystyle\sum_{i=1}^{N} \frac{K_i}{1+e^{\displaystyle -r_i \, (t-\tau_i)}} \, ,
\label{sumCN}
\end{equation}
which results, by differentiation, in the global incidence curve
\begin{equation}
    I(t) = \displaystyle\sum_{i=1}^{N} \displaystyle \frac{r_i \, K_i \, e^{\displaystyle -r_i \, (t-\tau_i)}}{ \left(\displaystyle 1+e^{\displaystyle -r_i \, (t-\tau_i)} \right)^2} \, .
    \label{INt}
\end{equation}

\subsubsection*{Calibration procedure}

The calibration process of each statistical model consists of identifying the underlying parameters with the aid of new deaths per day time series $(I_1, \cdots, I_n)$, that correspond to the discrete-time instants $(t_1, \cdots, t_n)$. This task requires the minimization of the \textit{Roots Mean Squared Error} (RMSE)
\begin{equation}
    RMSE = \sqrt{\dfrac{1}{n}\sum_{j = 1}^{n} \left(I_{j} - I(t_{j} \,) \right)^2} \,,
    \label{eq:RMSE}
\end{equation}
that measures the ``discrepancy'' between data values and the corresponding predictions given by the model that calculates the incidence $I(t)$. In practice, RMSE represents the weighted average of the square of the residuals generated by the calibrated model\cite{Wasserman2004,Hastie2009}.

The RMSE numerical minimization procedure employs a Trust Region algorithm\cite{Nocedal2006,Bonnans2009}, where bounds for the parameter value, as well as an initial guess, are specified. Such bounds are determined based on numerical experimentation followed by visual inspection of the fitted curves. To minimize the dependence of the fitted curve with the prescribed value for the initial guess, Monte Carlo simulation\cite{kroese2011,cunhajr2014p1355} is used, where random values for the initial guess (within the admissible region) are drawn and used to generate a fitting curve. Among all the obtained fittings, the one with the smallest RMSE is chosen.

To assess the quality of the statistical fit, this work also considers the coefficient of determination
\begin{equation}
    R^2 = 1 - \dfrac{\sum_{j = 1}^{n}\left( I_{j} - I(t_{j}) \,\right )^2 }{\sum_{j = 1}^{n} \left( I_{j} - \bar{I} \, \right)^2} \,,
    \label{Rsq}
\end{equation}
where the time series $(I_1, \cdots, I_n)$ average is given by
\begin{equation*}
    \bar{I} = \dfrac{1}{n}\sum_{j = 1}^{n} I_{j} \, .
\end{equation*}

The last metric describes how much of the total variance generated by the observed data is explained by the calibrated model. The value of $R^2$ varies between 0 and 1, so that the closer to 1 the more significant the model is, as the value of $R^2$ is the proportion of the original variability of the data that is explained\cite{Bishop2006,Hastie2009}.

\subsection{Model selection}

Different models can be fitted to the same dataset, choosing which is the most suitable requires the use of rational criteria to avoid potential bias. Thus, two information criteria are used here to choose the most suitable model following a rational criterion, which seeks to balance simplicity and predictability between the models. They are: (i) Akaike information criterion (AIC); (ii) Bayesian information criterion (BIC)\cite{Bishop2006,Hastie2009,Brunton2019}.

The AIC metric is given by
\begin{equation}
    AIC = -2 \, \log{ L(\hat{\theta})} + 2 \, p \,,
    \label{AIC}
\end{equation}
where $p$ is the number of model parameters, $\hat{\theta}$ the estimated parameters vector which maximizes the likelihood function $L(\theta)$. On the other hand, the BIC metric is defined as
\begin{equation}
    BIC = - 2 \ \log{ L(\hat{\theta})} + p \, \log{n} \,,
    \label{BIC}
\end{equation}
where $n$ is the number of observations. The model with the lowest AIC and BIC values is considered the best fitting model. Eventually these metrics can be combined with facts (documented in data) to assist in choosing the model.

\subsection{Outbreak start dates}

The estimate of the epidemic outbreak start date is not done in a punctual way, because an exact date in a context full of uncertainties such as an epidemic outbreak is a fragile estimate, and meaningless from any modeling point of view that makes sense. Thus, an interval estimate is considered in this work, where the upper limit of an admissible interval for the beginning of the outbreak is estimated with the help of the confidence band that encompasses the uncertainty of the statistical model. In this way, after the construction of a statistical model that fits the data, a 95\% confidence interval is obtained around the model's response curve. Such a prediction interval\cite{Wasserman2004}, with $(1-\alpha) \times 100\%$ confidence, is defined by
\begin{equation}
    I(t) \pm z_{\alpha/2} \, \xi_{n} \,,
    \label{ICzj2}
\end{equation}
where $z_{\alpha/2}$ is the quantile of the Student-$\text{t}$ distribution with $n$ degrees of freedom and statistical significance $\alpha = 0.05$, while
\begin{equation}
    \xi_{n} = S \, \sqrt{1 + \frac{1}{n} \, \frac{\sum_{j = 1}^{n}(I_{j} - I(t)\,)^2}{\sum_{j = 1}^{n}(I_{j} - \bar{I} \,)^2}} \,,
    \label{ICzj2}
\end{equation}
with
\begin{equation}
    S = \frac{\sum_{j=1}^n (I(t_j)-I_j)^2}{n-2} \, .
\end{equation}

The intersection between the boundaries of this envelope and the time axis provides a range of possible dates for the start of the outbreak under investigation so that the rightmost point of this intersection is assumed to be an upper bound for the outbreak start date.

% --------------------------------------------------------------

% --------------------------------------------------------------
\section{Results and discussion}
\label{sec:Results and discussion}

In this section, our objective is to obtain an interval estimation for the starting dates of each COVID-19 outbreak in Rio de Janeiro city.  The start date in this context is understood to be the day on which a certain level of prevalence is achieved so that new infection events occur every day from that point forward. This starting date may be related to the introduction of a novel viral strain in the community or to a key event that initiated a new chain of contagion.

\subsection*{The starting of SARS-CoV-2 community transmission}

The data shown in Fig.~\ref{fig:fig3} allows us to observe the existence of several waves of contagion in Rio de Janeiro city, which resulted in 6 waves of deaths, four big explosions, and two small boosts. In this first analysis, only the first of these waves is considered.

Once the exact date where one wave ends and another begins is extremely uncertain, makes more sense to speak about this event in an interval sense, a plausible period where that date is contained. In this way, by visual inspection of Fig.~\ref{fig:fig3}, it is verified that the first wave of deaths starts between the middle of March and the beginning of April, and ends by the end of June or the beginning of July 2020. Between March and April 2020, the epidemic surveillance system in Rio de Janeiro was still adapting to the pandemic, testing to determine if death from respiratory disease was due to COVID-19 was not yet mandatory, so that underreporting in this period (certainly above average) can bias the estimate of the outbreak start date. In May 2020, on the other side, the COVID-19 death tracking system was better developed, so the first wave phase records should not be so biased. Thus, for the sake of minimizing bias during statistical model calibration, the training data considered here uses the time series of new deaths per day between May 1st and July 1st.

Table~\ref{tab1} shows the values of the parameters found in the calibration process for a single wave model, the estimated date for the epidemic outbreak start, as well as the quality metrics of the fitting and information criteria, for several scenarios of the parameter $\tau$, which represents the instant of inflection of the logistic model (the peak of the incidence curve). These multiple scenarios are considered because the evolution of a patient who dies, between the time of infection and death, has a variable duration. Therefore, considering a range of possible dates around the peak of the data (April 28, day 127 of the time series) is a strategy to reduce the influence of this uncertainty on the inference process. By generating a family of possible curves (ensemble of models), the adopted information criteria (AIC and BIC) and the known facts about the outbreak will indicate which of those curves are a plausible representation for the wave of deaths.

\begin{table*}
\centering
\caption{{\bf Estimated parameters, the respective confidence intervals, fitting metrics, and information criteria measures for the statistical models used to represent the first wave of deaths by COVID-19 in Rio de Janeiro (for several scenarios of peak day). 
}}
\label{tab1}
\vspace{2mm}
\begin{tabular}{c|c|c|c|c|cc|cc}
\toprule
$\tau$ & $K \times 10^3$ & $r \times 10^{-3} $ & deaths start date & cases start date & RMSE & R$^2$ & AIC & BIC\\
(day)	   & (people)               & (day$^{-1}$)             & upper bound      & upper bound      &         &            &        &      \\
\midrule
119	   & 9.70 (9.14, 10.26) & 53 (49, 57) & Mar 13, 2020 & Mar 02, 2020 & 13.8 & 0.86 & 3.99 & 8.24\\
120	   & 9.70 (9.18, 10.21) & 53 (49, 57) & Mar 12, 2020 & Mar 01, 2020 & 12.7 & 0.89 & 3.99 & 8.24\\
121	   & 9.70 (9.21, 10.18) & 53 (49, 56) & Mar 11, 2020 & Feb 29, 2020 & 11.8 & 0.90 & 3.99 & 8.24\\
122	   & 9.70 (9.23, 10.16) & 53 (50, 56) & Mar 11, 2020 & Feb 29, 2020 & 11.1 & 0.91 & 4.19 & 8.44\\
123	   & 9.70 (9.25, 10.15) & 52 (49, 55) & Mar 11, 2020 & Feb 29, 2020 & 10.7 & 0.92 & 4.04 & 8.30\\
125	   & 9.28 (8.84, ~9.73) & 55 (51, 57) & Mar 15, 2020 & Mar 04, 2020 & 10.8 & 0.92 & 3.98 & 8.24\\
127	   & 8.72 (8.28, ~9.15) & 57 (53, 60) & Mar 20, 2020 & Mar 09, 2020 & 11.0 & 0.91 & 4.01 & 8.27\\
129	   & 8.18 (7.74, ~8.62) & 60 (56, 64) & Mar 26, 2020 & Mar 15, 2020 & 11.4 & 0.91 & 4.00 & 8.25\\
131	   & 7.68 (7.22, ~8.13) & 63 (59, 68) & Mar 31, 2020 & Mar 20, 2020 & 12.2 & 0.89 & 4.00 & 8.26\\
\bottomrule
\end{tabular}
\end{table*}

Note that for each model corresponds to a date, which is the estimate of the upper limit for the beginning of the wave of deaths. Due to the correlation between cases and deaths, there is a temporal shift (with a certain probability distribution) between this limit date on the curve of deaths and its counterpart on the curve of cases. Hawryluk et al.\cite{Hawryluk2020} points out that the COVID-19 average time from symptom onset to death in Brazil, during the first wave, was around 15.2 (11.2, 17.8) days, so that a conservative estimate for the upper bound for the date of onset of cases (fifth column of Table~\ref{tab1}) can be obtained by subtracting 11 days from the date estimated by the curve of deaths (fourth column of Table~\ref{tab1}).

\begin{figure*}[ht!]
    \centering
    \includegraphics[scale = 0.3]{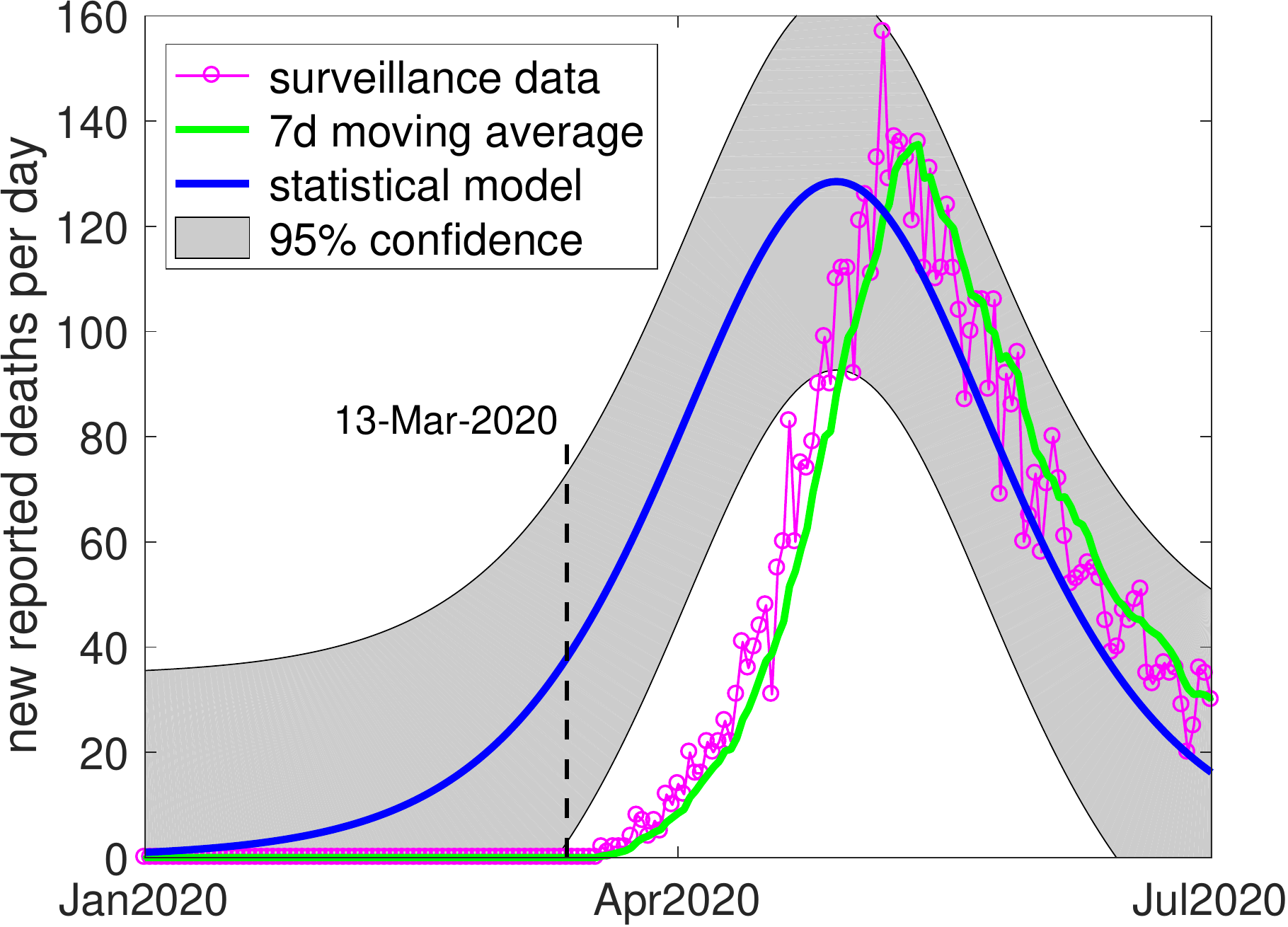}
    \includegraphics[scale = 0.3]{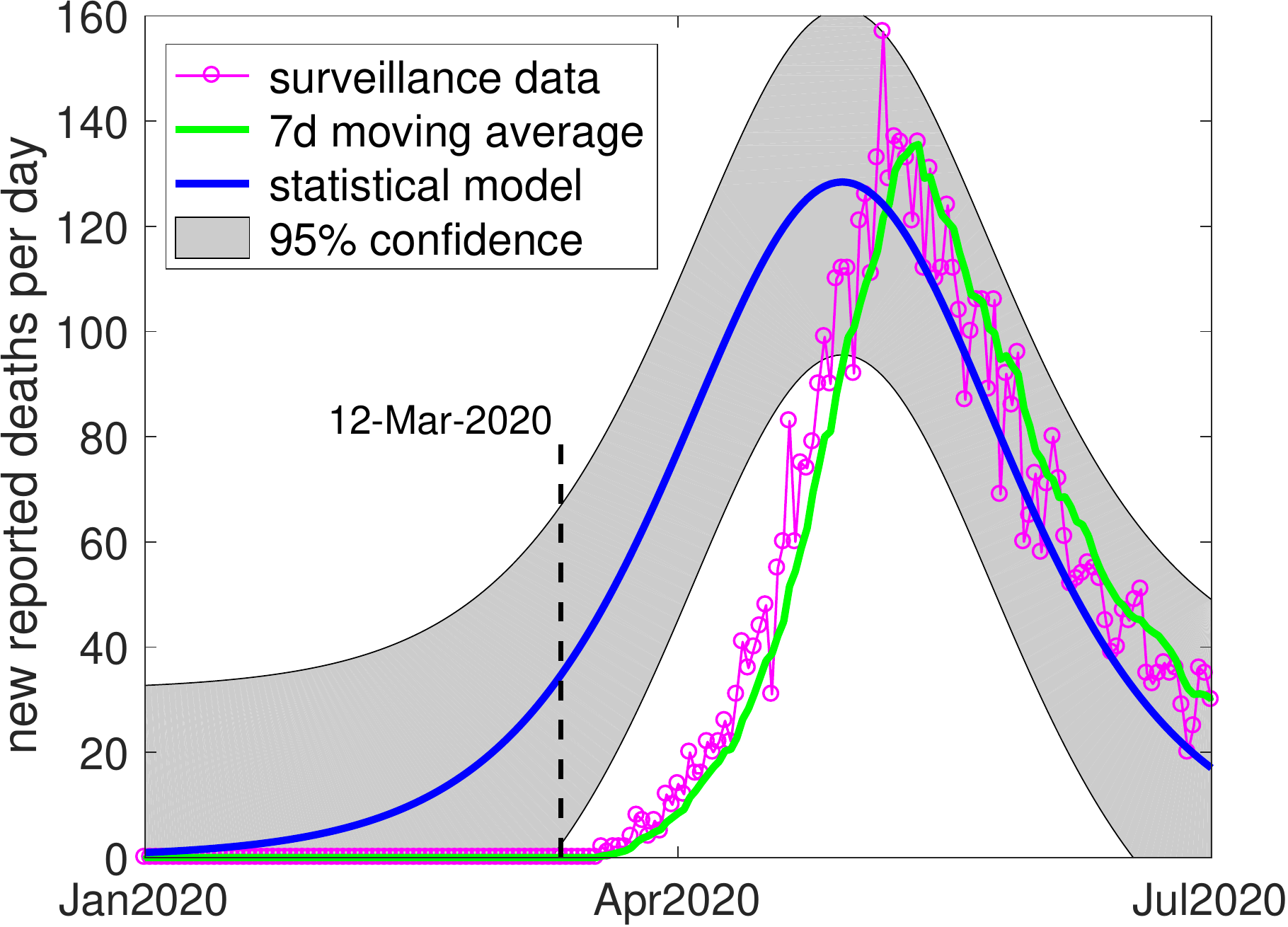}
    \includegraphics[scale = 0.3]{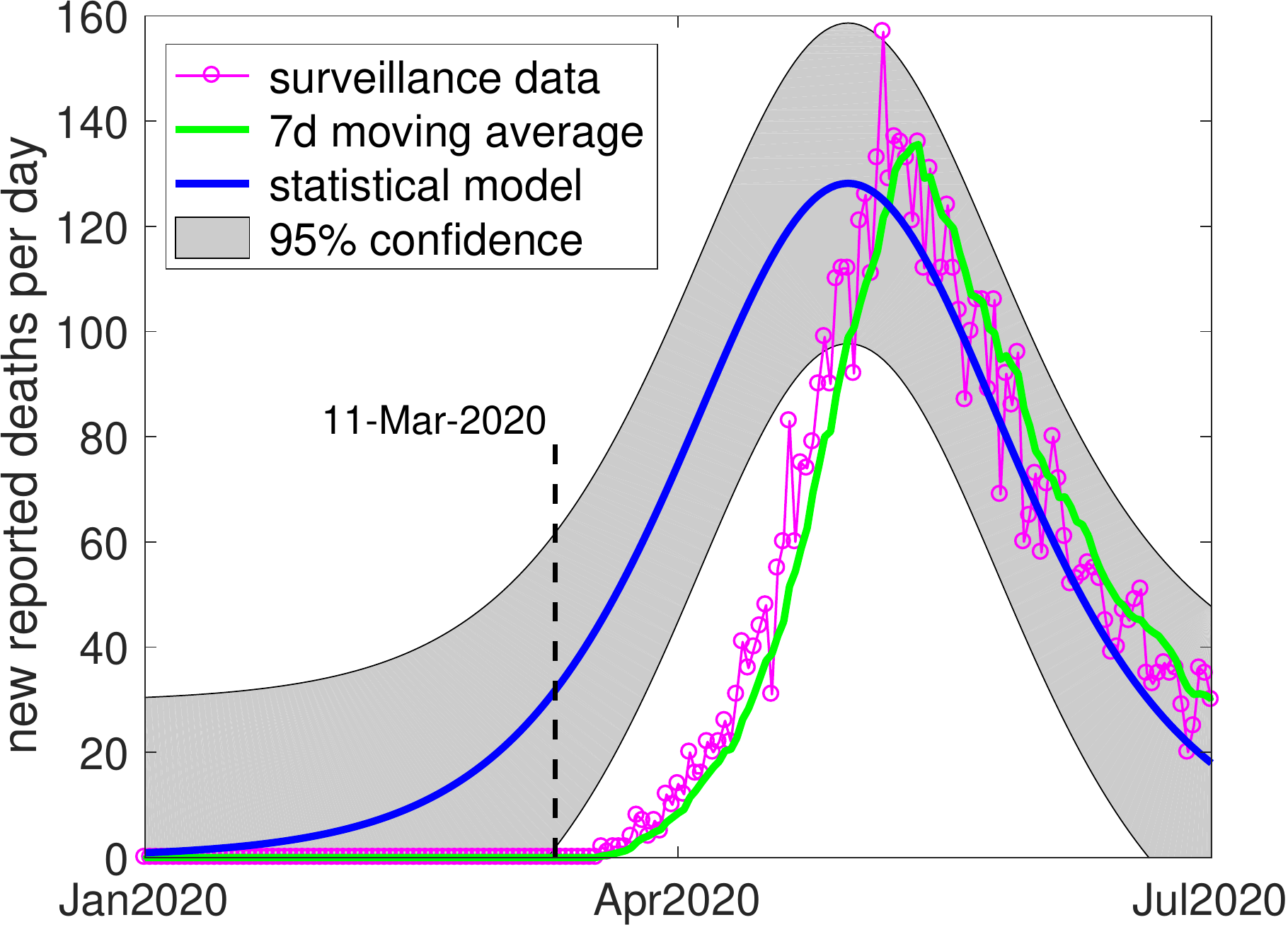}\\
	\vspace{5mm}    
	
    \includegraphics[scale = 0.3]{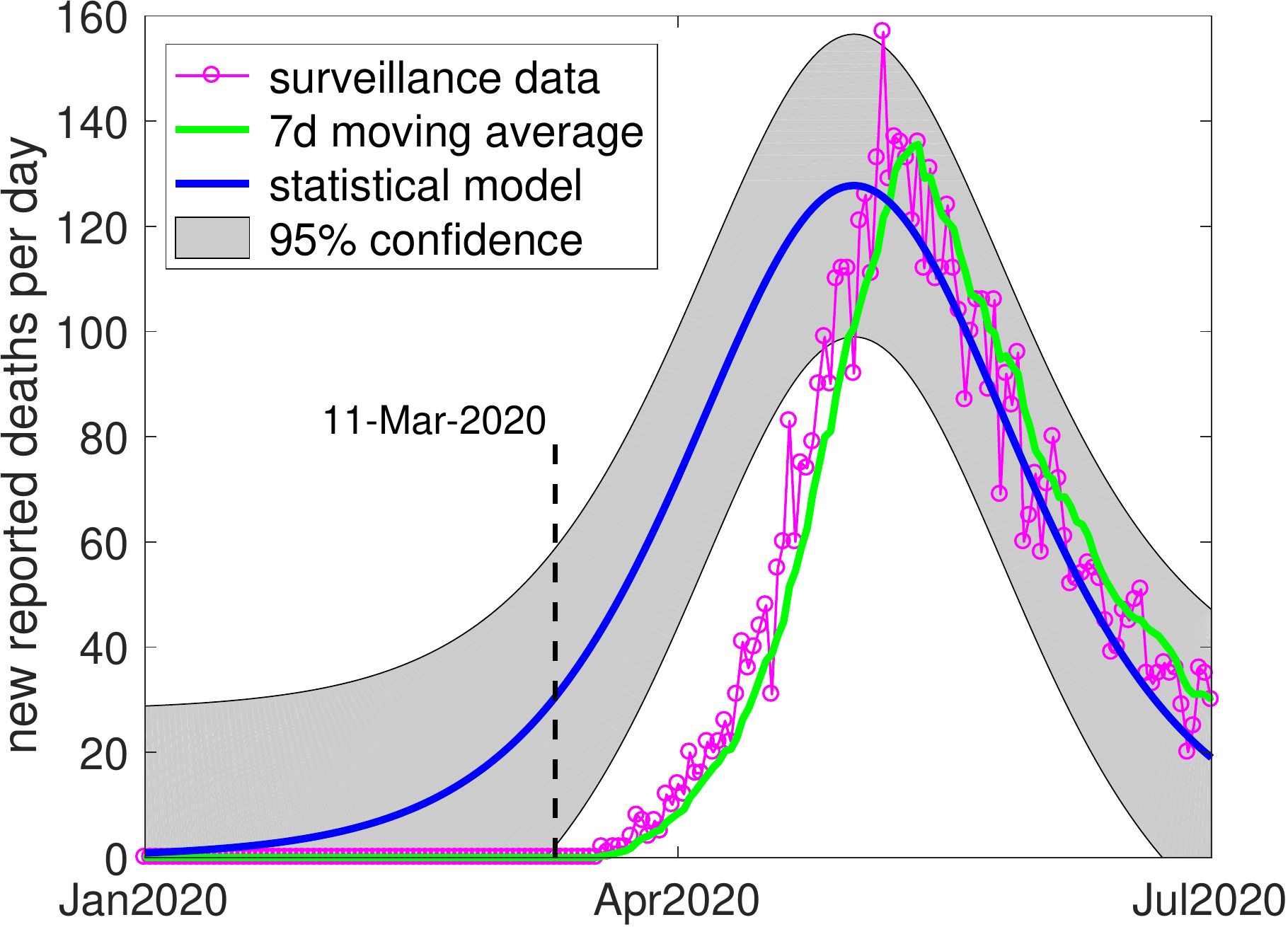}
    \includegraphics[scale = 0.3]{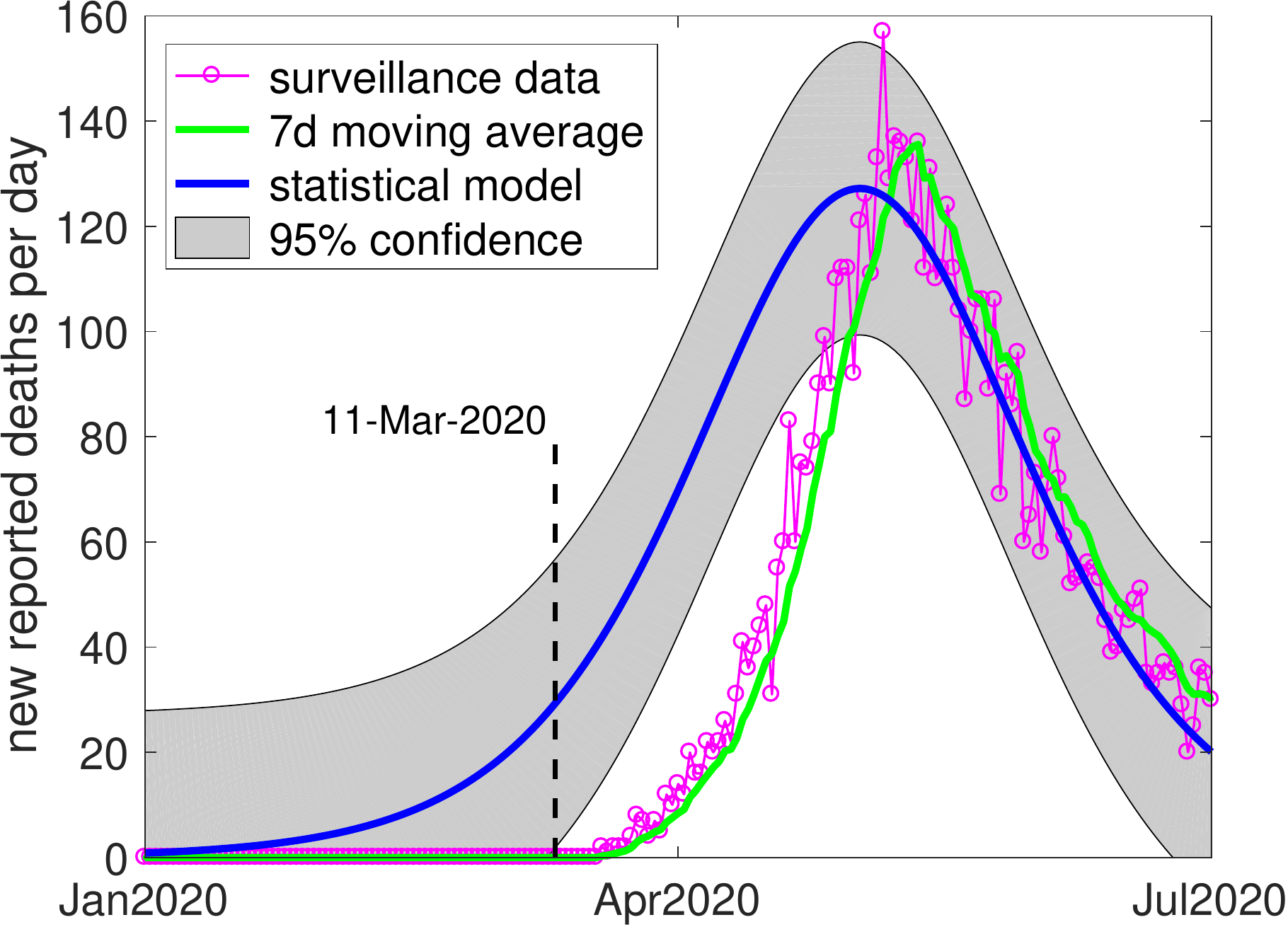}
    \includegraphics[scale = 0.3]{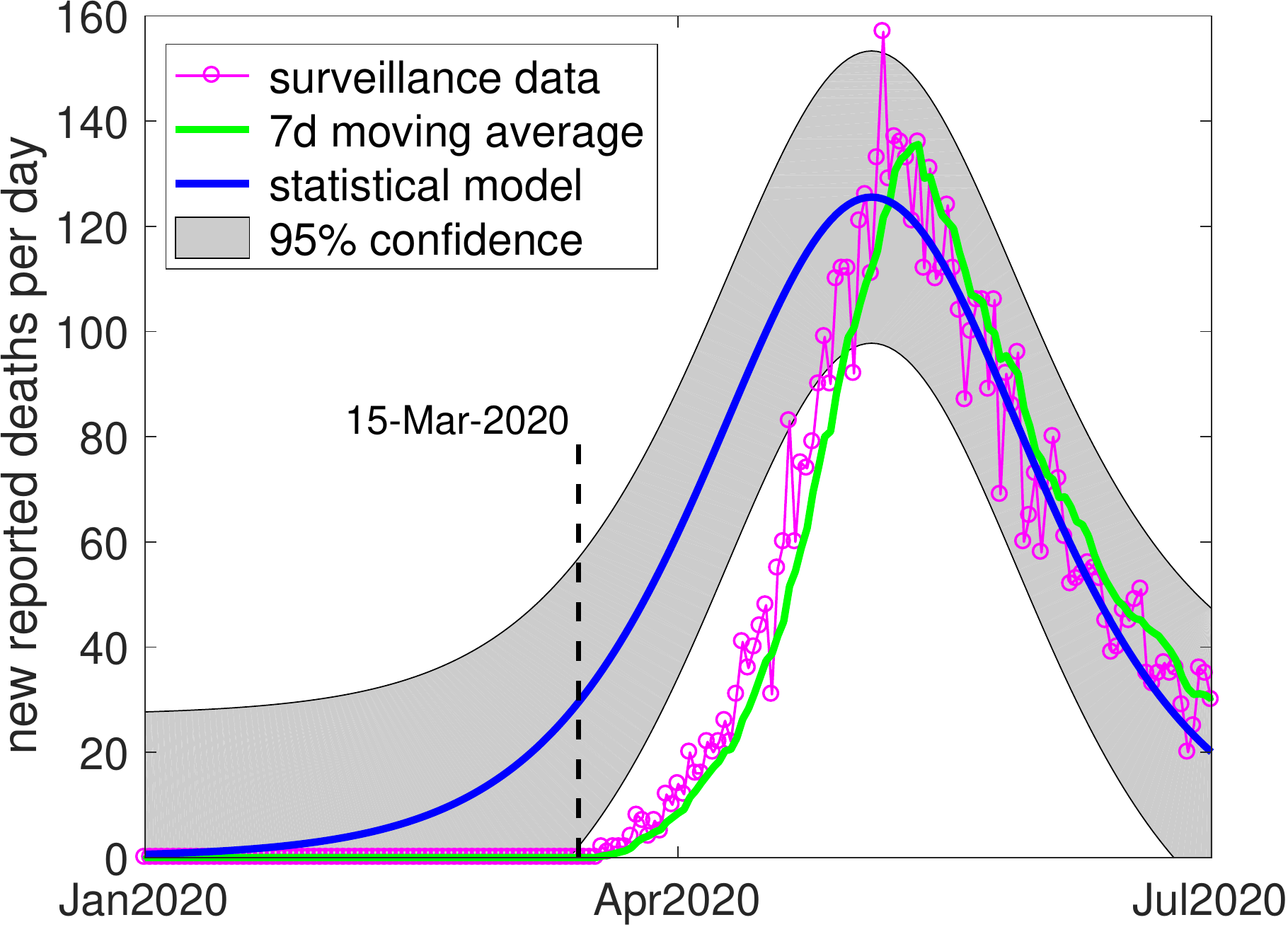}\\
	\vspace{5mm}    
    
    \includegraphics[scale = 0.3]{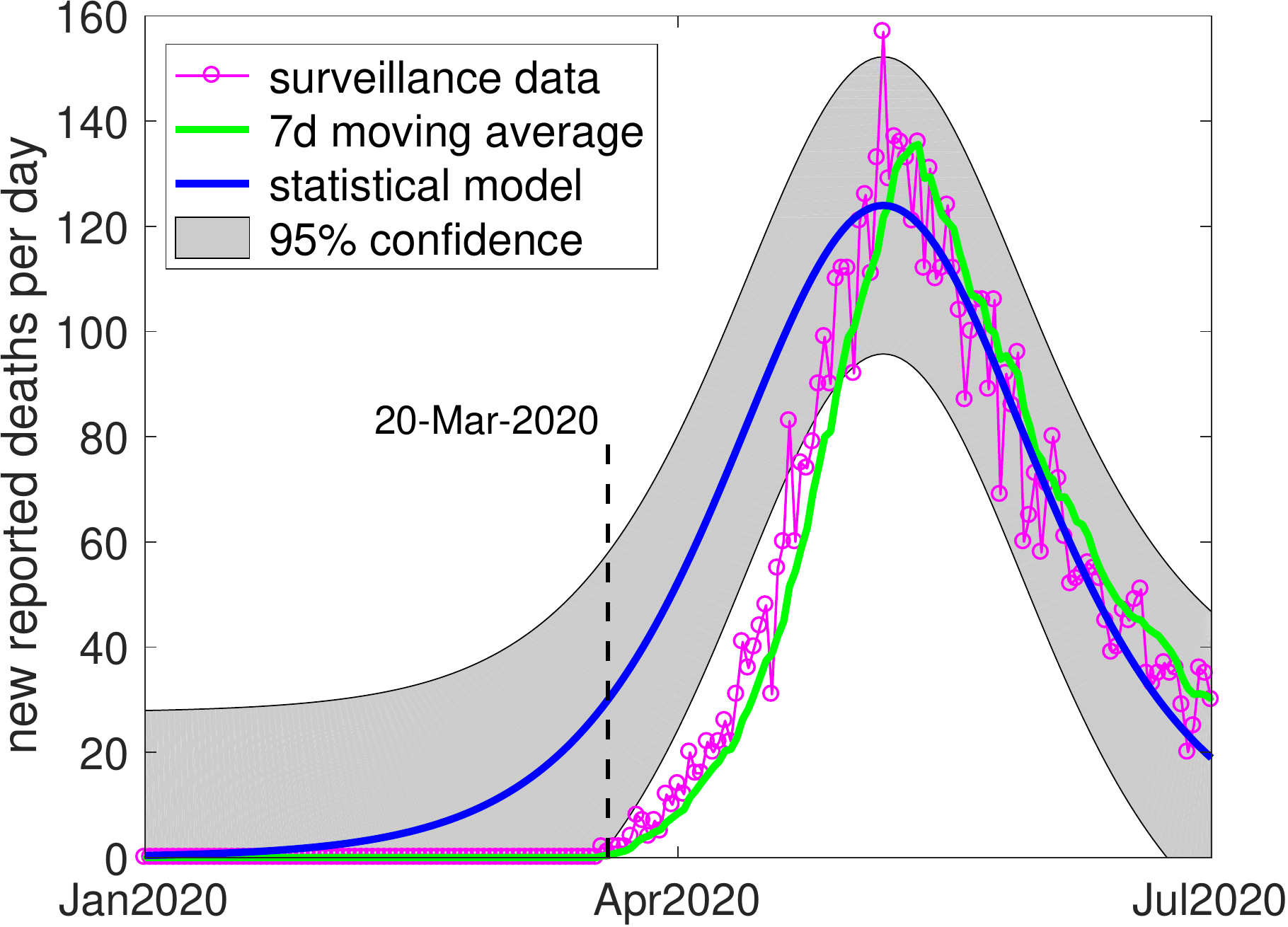}
    \includegraphics[scale = 0.3]{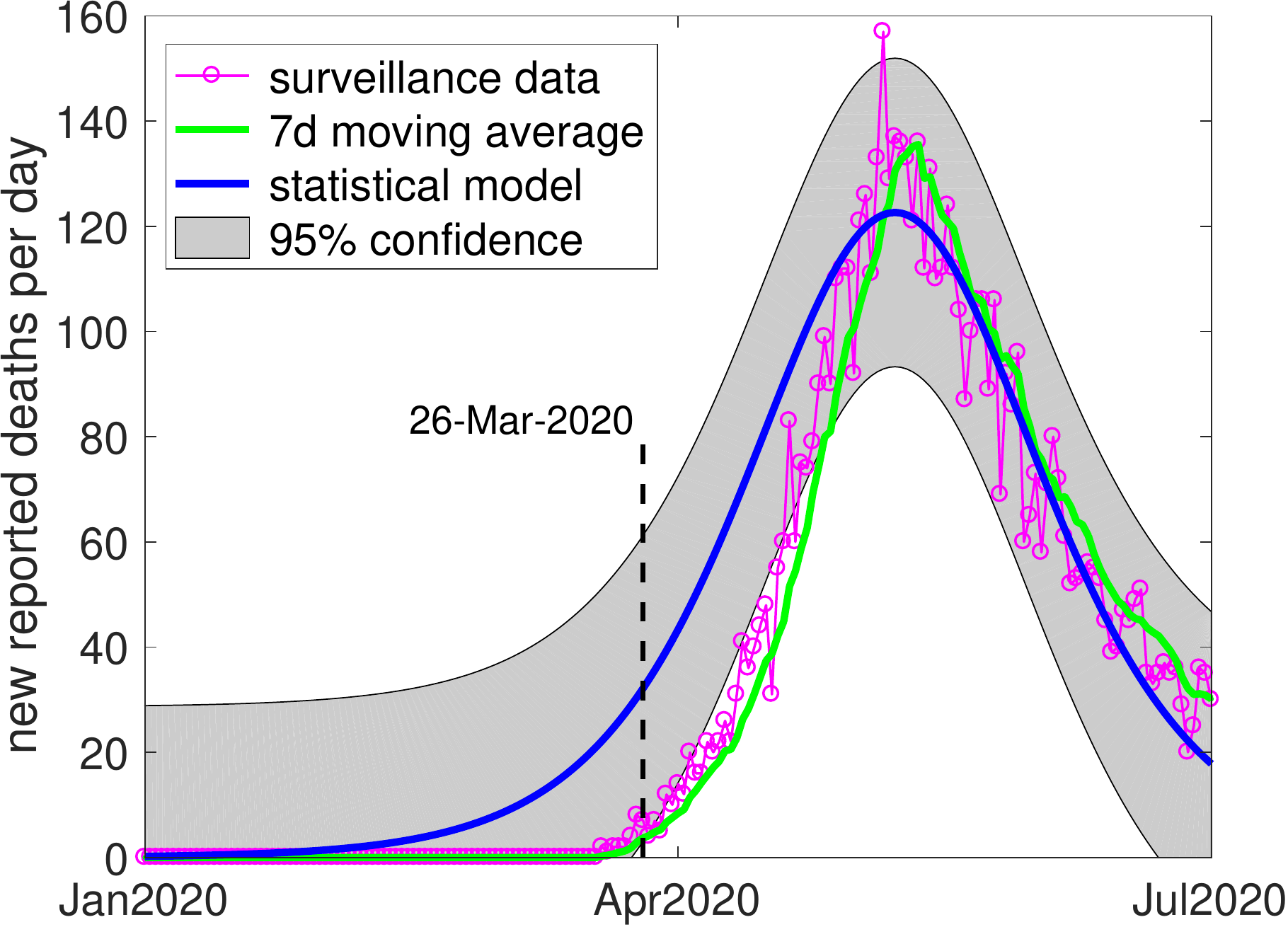}
    \includegraphics[scale = 0.3]{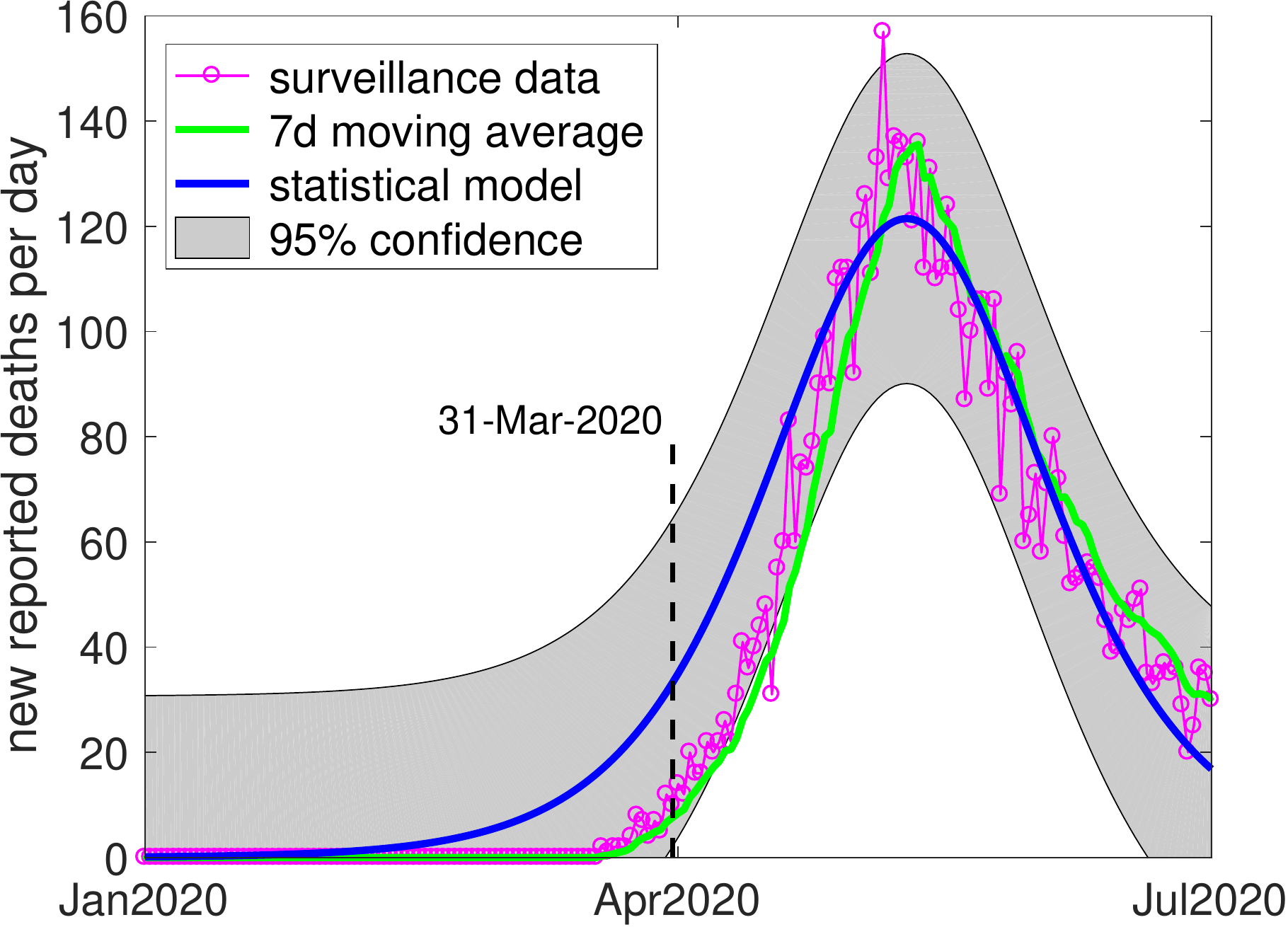}
    \caption{Logistic curves fitted to the time series associated with the incidence of deaths during the first wave of COVID-19 in Rio de Janeiro city, and the corresponding prediction band, for several scenarios of peak day: 119 (top left), 120 (top center), 121 (top right), 122 (middle left), 123 (middle center), 125 (middle right), 127 (bottom left), 129 (bottom center), 131 (bottom right). The dashed vertical lines indicate the upper bound estimation for the starting date of the first epidemic wave. Data used in the model calibration: May 1st to July 1st, 2020.}
    \label{fig:fig5}
\end{figure*}

\begin{figure*}[ht!]
    \centering
    \includegraphics[scale = 0.3]{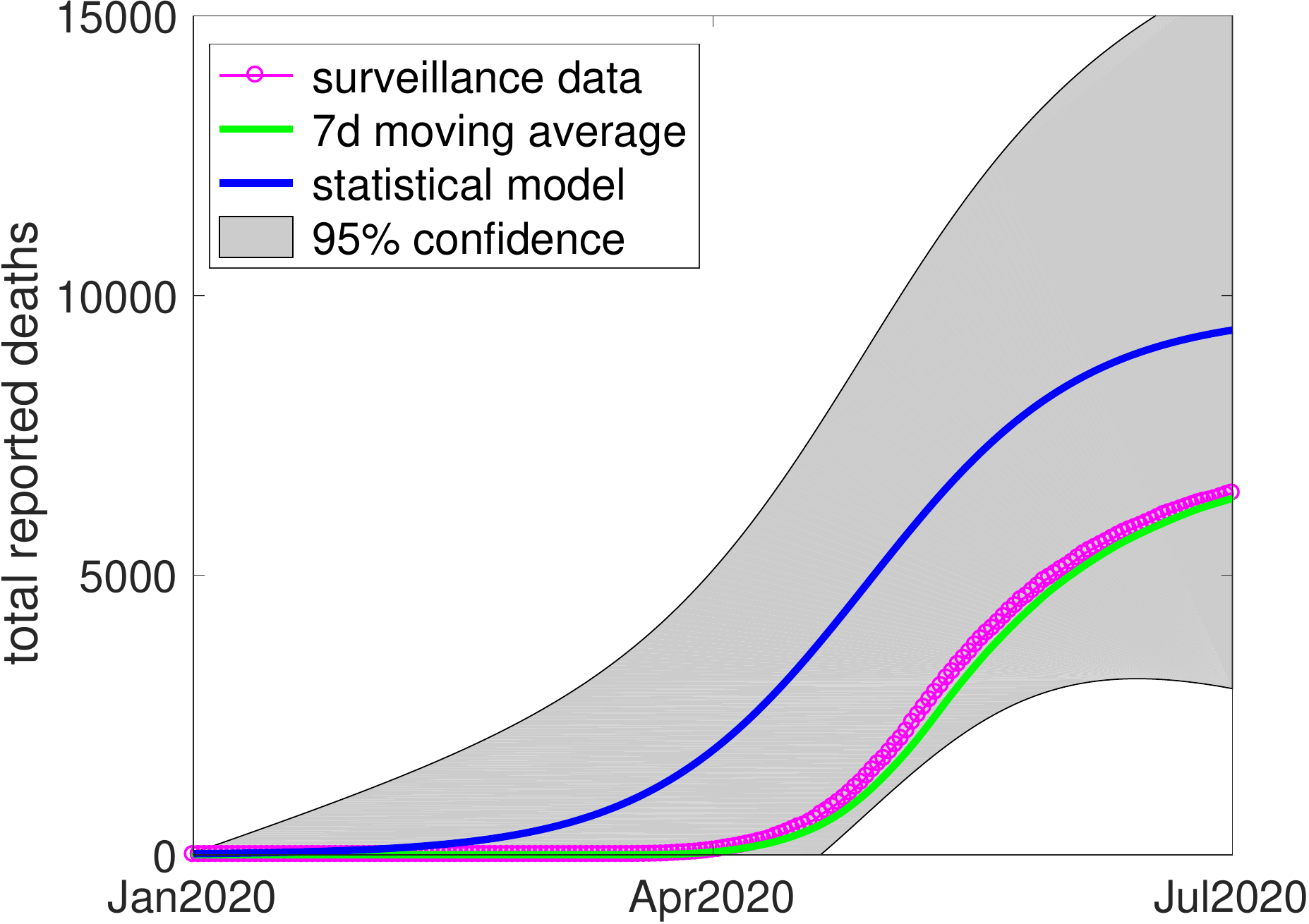}
    \includegraphics[scale = 0.3]{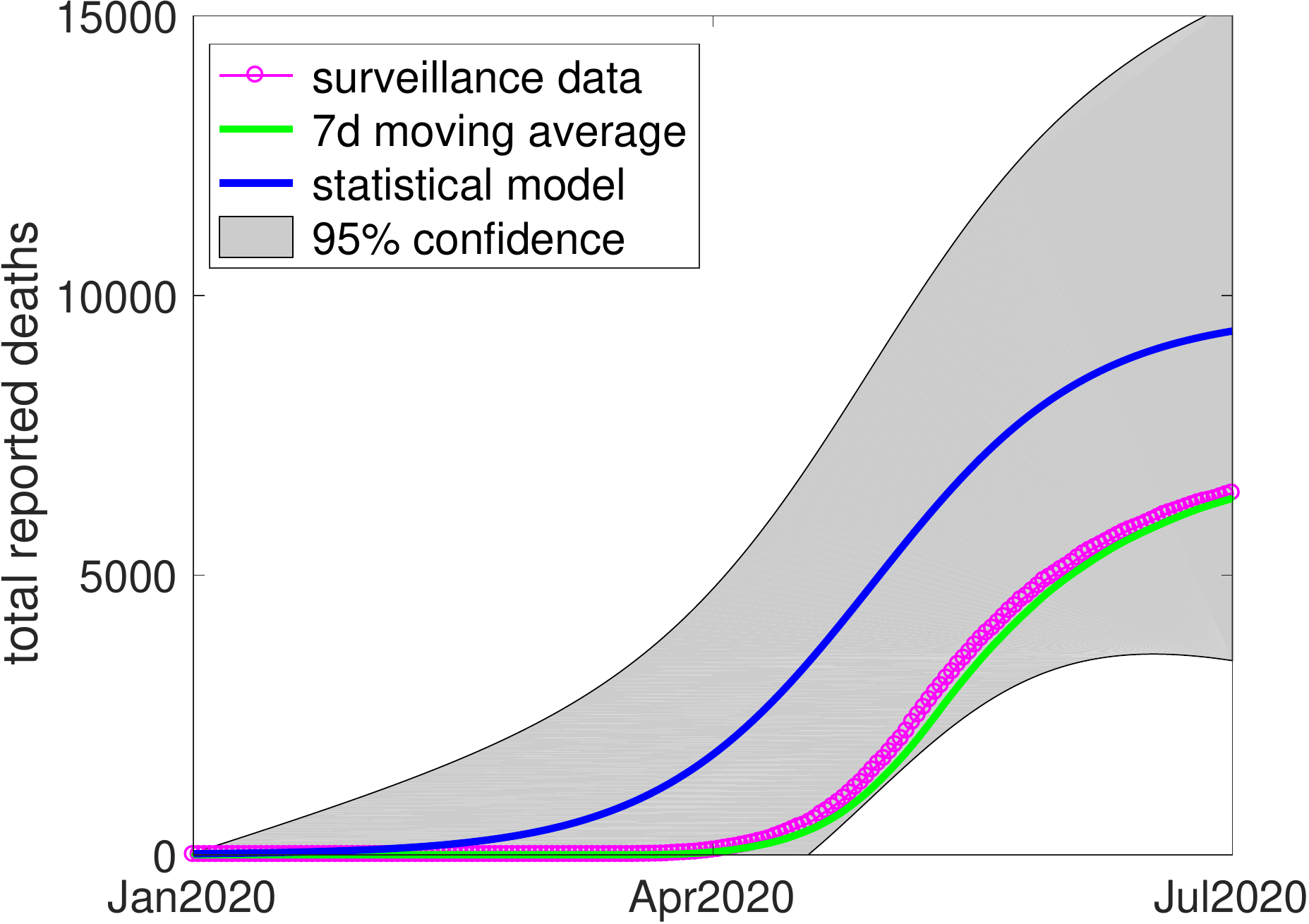}
    \includegraphics[scale = 0.3]{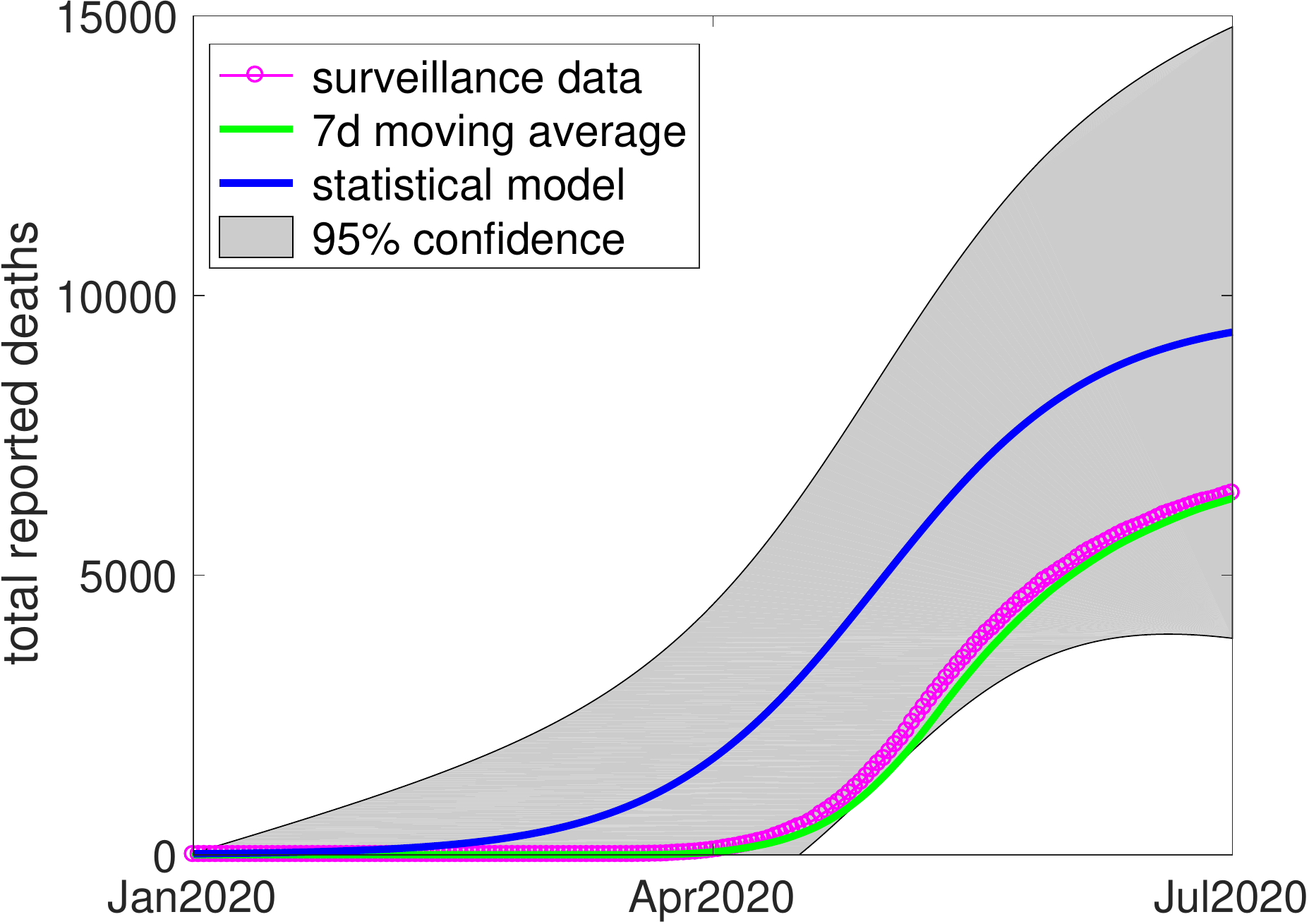}\\
	\vspace{5mm}    
    
    \includegraphics[scale = 0.3]{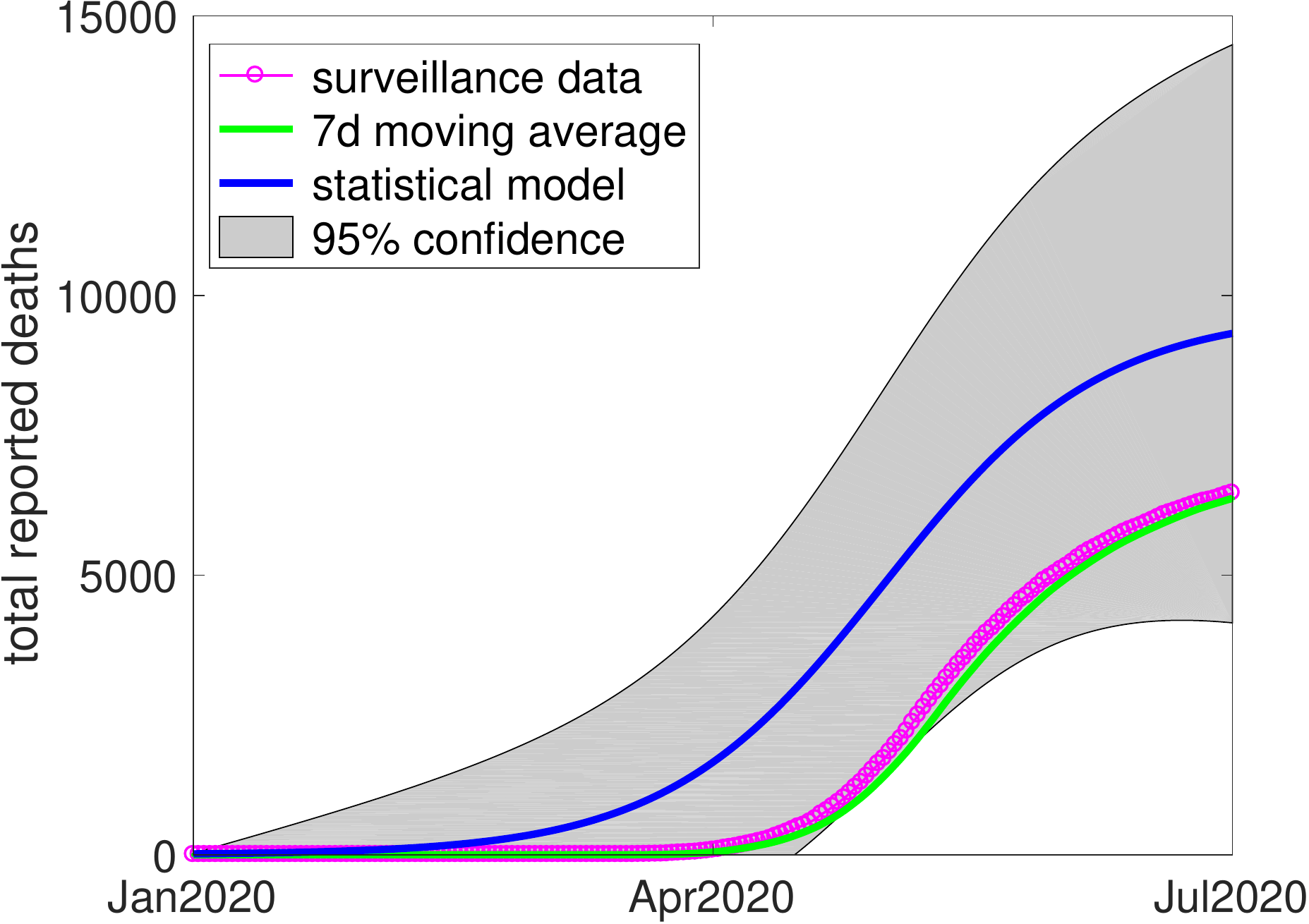}
    \includegraphics[scale = 0.3]{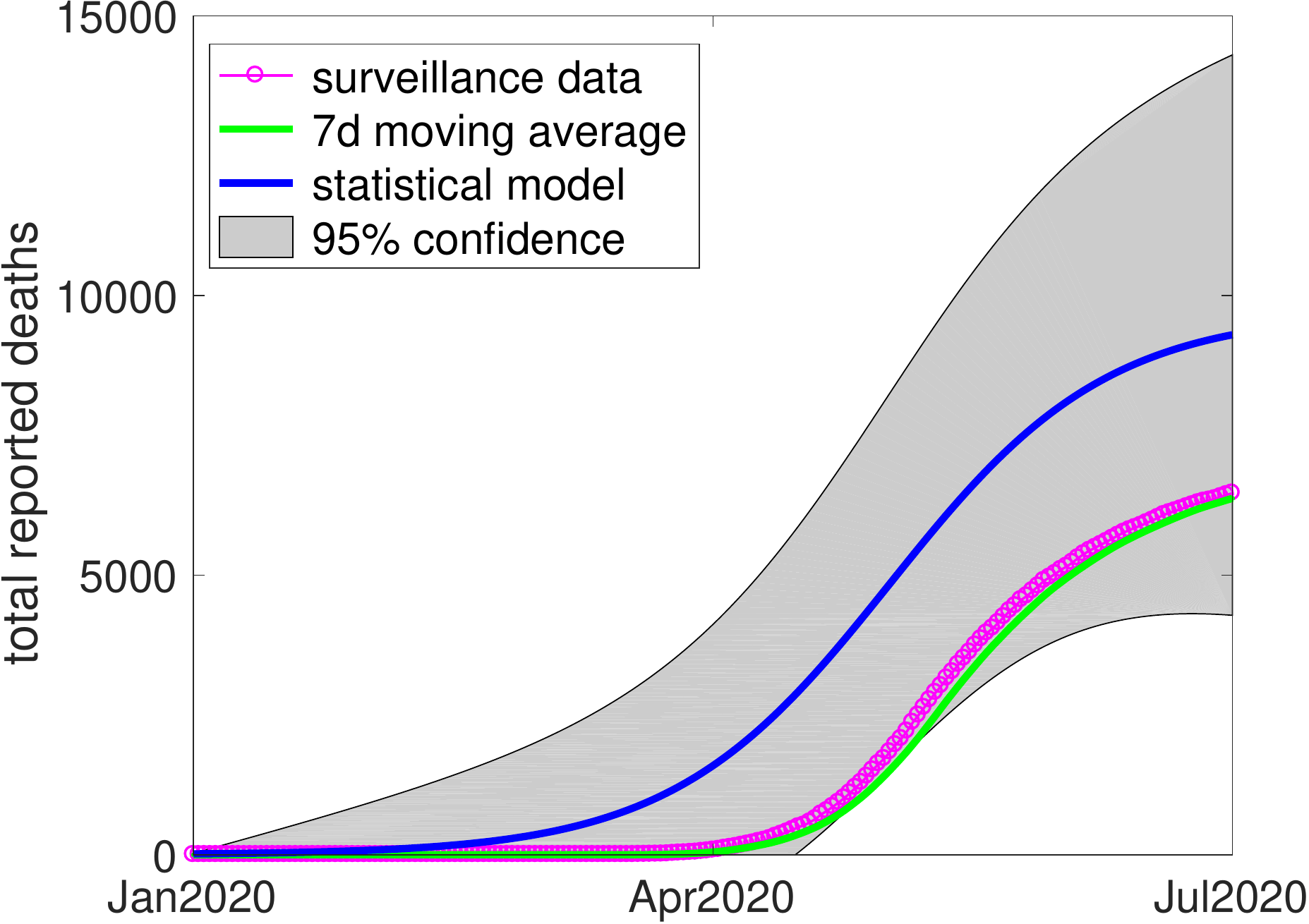}
    \includegraphics[scale = 0.3]{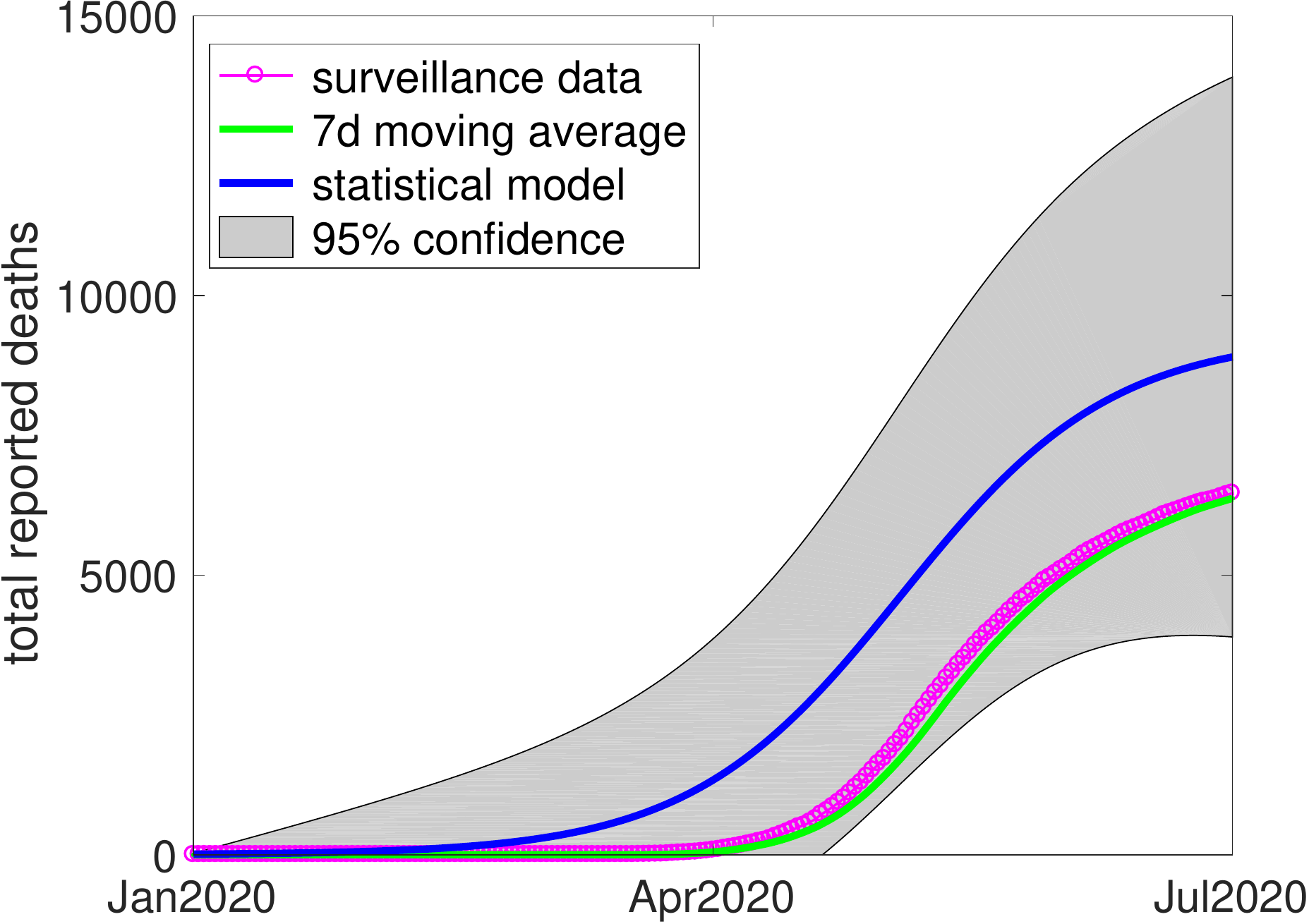}\\
	\vspace{5mm}    
    
    \includegraphics[scale = 0.3]{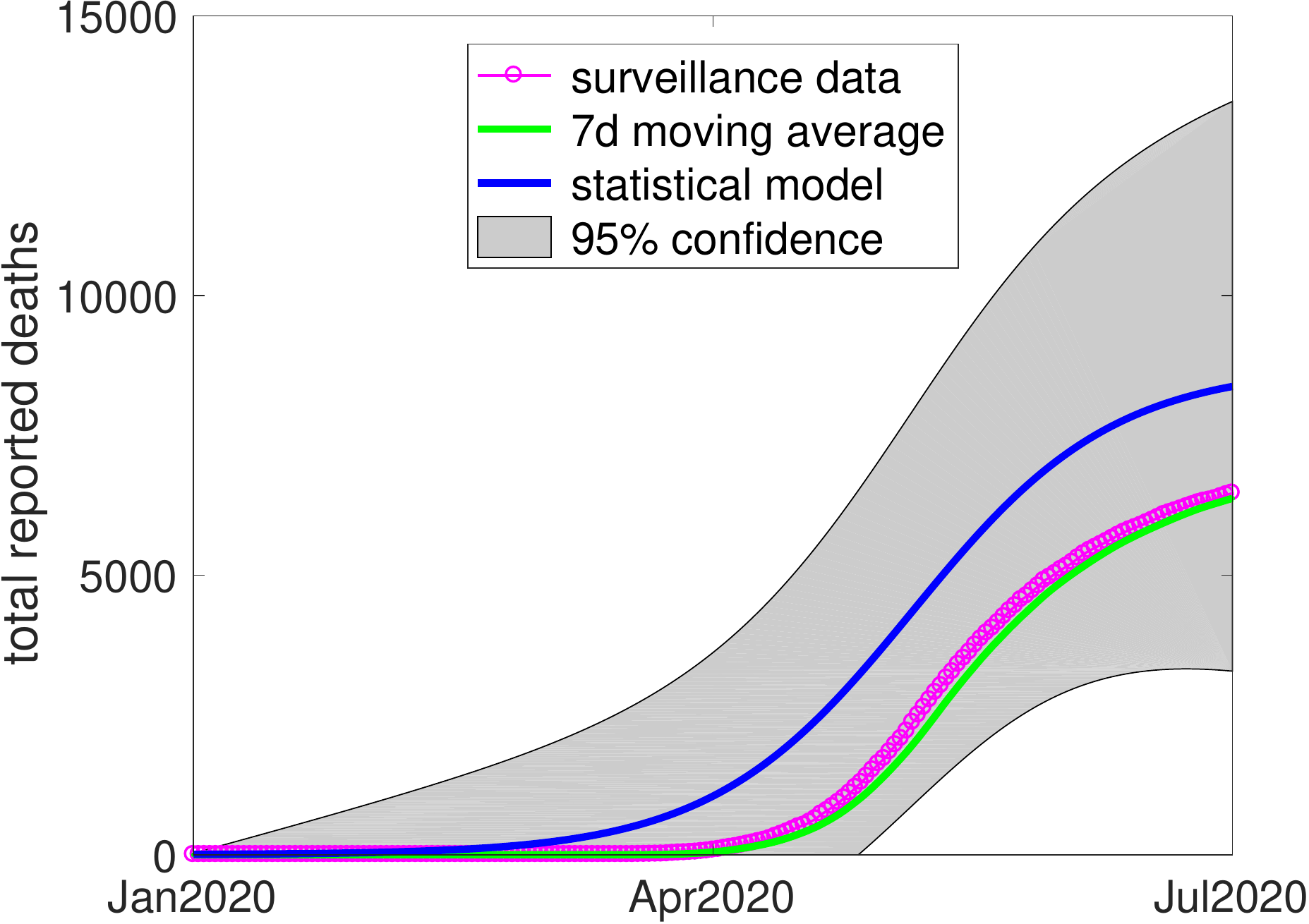}
    \includegraphics[scale = 0.3]{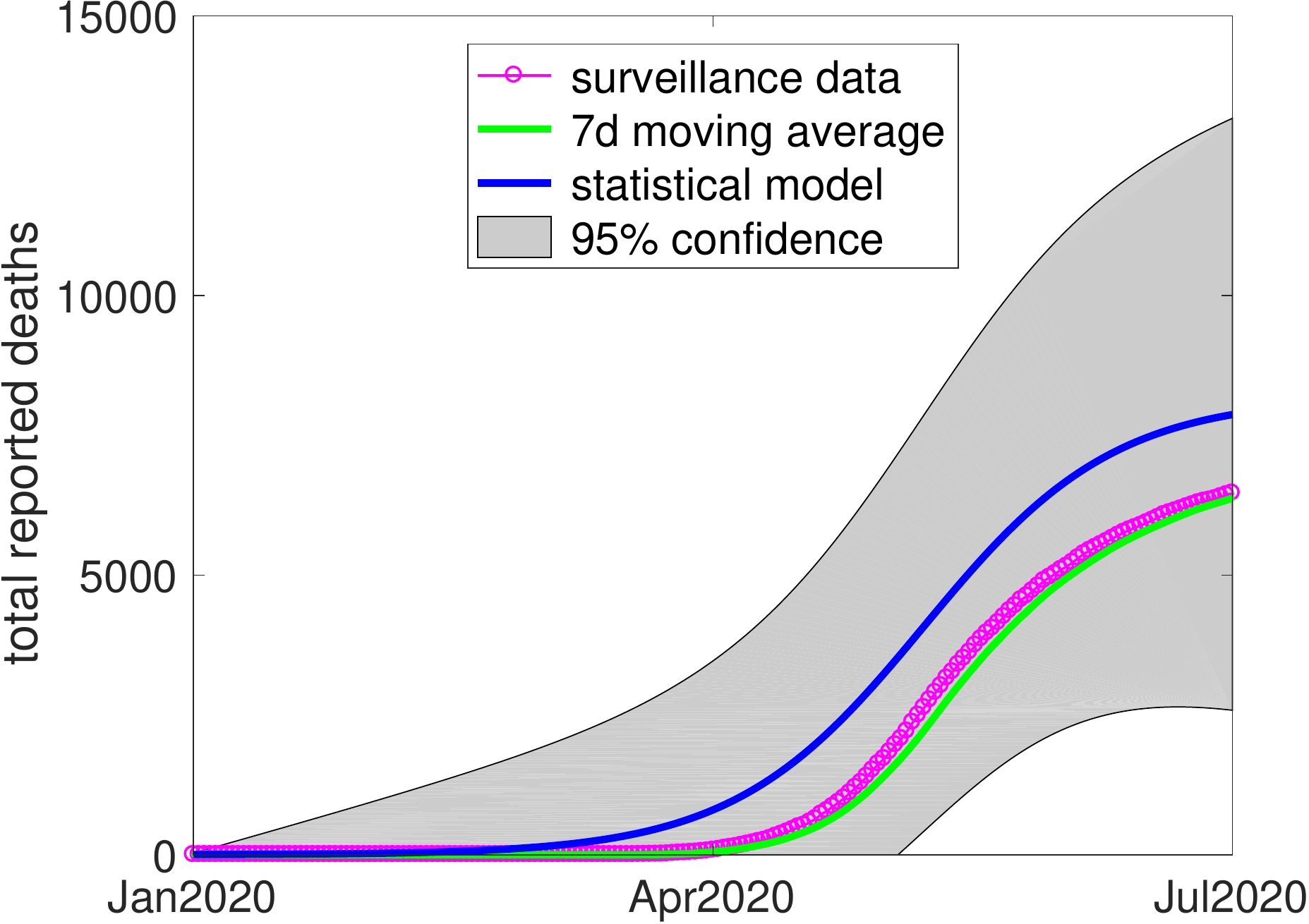}
    \includegraphics[scale = 0.3]{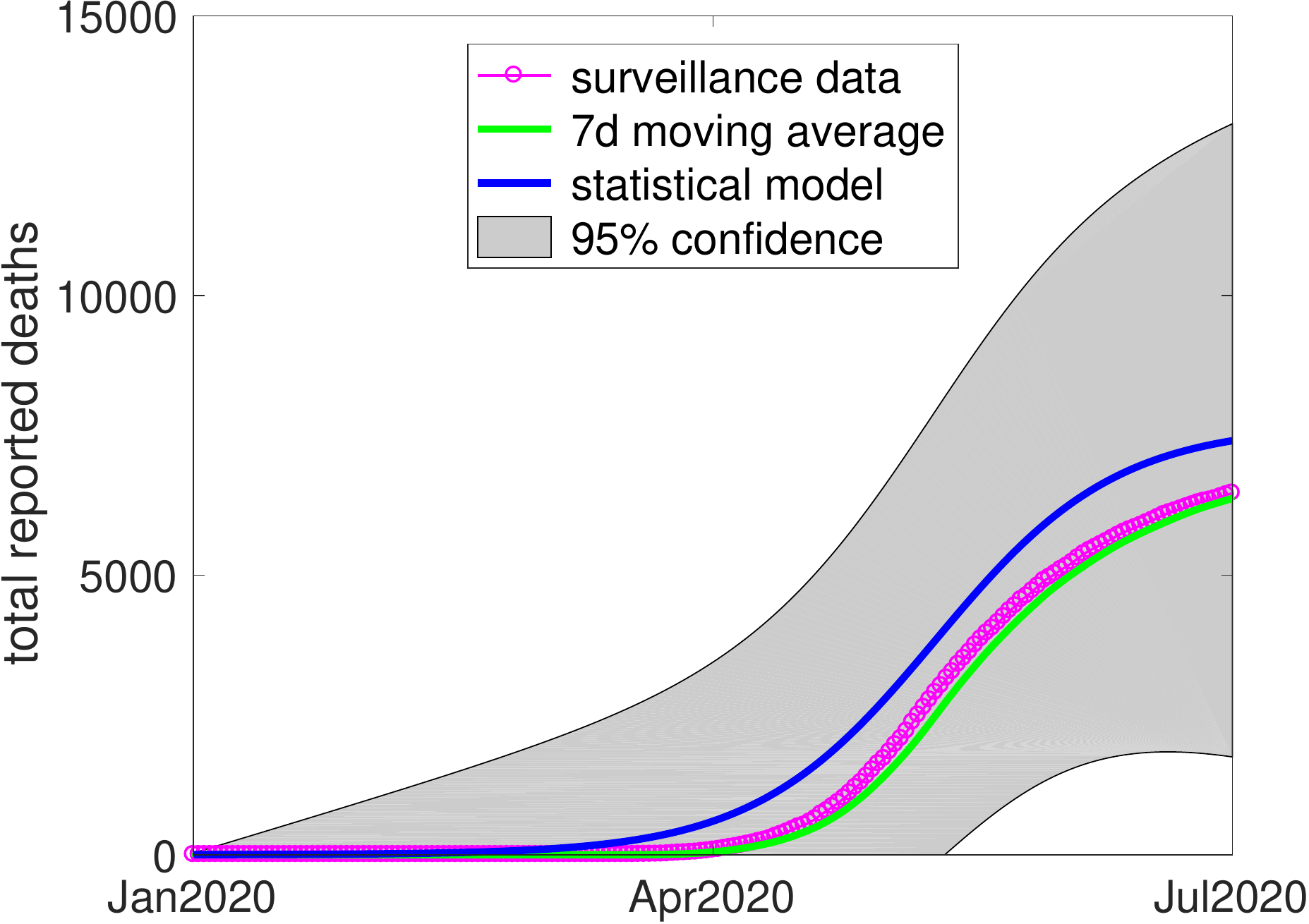}\\
    \caption{Logistic curves fitted to the time series associated with the prevalence of deaths during the first wave of COVID-19 in Rio de Janeiro city, and the corresponding prediction band, for several scenarios of peak day: 119 (top left), 120 (top center), 121 (top right), 122 (middle left), 123 (middle center), 125 (middle right), 127 (bottom left), 129 (bottom center), 131 (bottom right). Data used in the model calibration: May 1st to July 1st, 2020.}
    \label{fig:fig6}
\end{figure*}

Furthermore, the fitting metrics of Table~\ref{tab1} suggest that all the calibrated statistical models have good adherence to the data, which can be visually confirmed in Fig.~\ref{fig:fig5}, which shows 9 candidate models for the single wave incidence curve, each one corresponding to a different value for $\tau$.  The corresponding prevalence curves can be seen in the sequence, in Fig.~\ref{fig:fig6}. In these figures, the raw epidemic data is represented by magenta dots, the 7 days moving average by a thin green line, the fitted epidemic curve is given by a thick blue line, and the 95\% confidence band is shown in gray.

In qualitative terms, all the candidate models provide a good description of the descending part of the wave, and they all more or less agree on the inference of the "hidden part" of the wave, i.e., that part of the outbreak which was not captured by the epidemic surveillance system on the early stages of the epidemic process. It is also noted that models with a lower value of $\tau$ (inflection point furthest from the peak of data) tend to estimate underreporting in death records more conservatively, as their prevalence curves are more detached from the observations (known to be underreported). However, despite the qualitative similarity between all candidate models, from a quantitative point of view, differences exist and materialize themselves in the estimation of the upper bound for the starting dates. These quantitative differences in date predictions are compared with known information about the onset of community transmission, to eliminate models that provide predictions inconsistent with observations. This can be seen as a complementary step in the model selection procedure, to delineate the most plausible model(s) for the available data.

The surveillance data used to construct the Fig.~\ref{fig:fig3} presents sparse notifications for cases in January and February, and no null records from March 1st. This is a direct indication that at the beginning of March there was already community transmission. Thus, by inconsistency with direct evidence, we will eliminate all models that point to a date greater than March 1st as the upper bound for the wave of cases start date, which leaves us with four candidate models, corresponding to  $\tau \in \{120, 121, 122, 123 \}$.

Among these four models, $\tau = 120$ and $121$ have the lowest values for AIC and BIC, being in principle the natural choices for the most representative model. And between these two candidate models, $\tau =121$ has the lowest RMSE and highest $R^2$, so it is the optimal choice in light of fitting metrics and model selection criteria. This model indicates \underline{\emph{February 29, 2020}} as an upper bound for the first wave of cases starting date.

\subsection*{Consistency on the inference of first wave starting date}

It is very clear that the comparison with the evidence of the beginning of community transmission provides a good filter to discard models that overestimate the desired date, but does not provide information about the other models. This role is in charge of the model selection criteria, which show a marginally small difference in their numerical values for all the remnants candidate models, which may indicate that all the models under evaluation are more or less equally capable of describing the first epidemic wave. The fitting metrics RMSE and $R^2$ can be used as a supplement, in this case, the drawback is that, in the case of overfitting, they may provide very poor indicators of the model quality. Thus, in a scenario of metrics, as indicated in Table~\ref{tab1}, additional considerations to indicate the consistency of the statistical fit are desirable.

Following this idea, a combination (arithmetic mean) of the forecast obtained by the four candidate models is used as an estimator for the starting date of cases, providing again \underline{\emph{February 29, 2020}} as a possible upper limit for the beginning of community transmission of COVID-19 in the city of Rio de Janeiro. Once this is an interval estimate, it is possible (and probable) that the threshold of dozens of cases in the prevalence of the disease has occurred sometime in February, instead of March as current data suggest.

To strengthen this conclusion we use a hypothesis test, where the null and alternative hypotheses adopted are $$H_0: \mbox{cases start date} \geq \mbox{March 1, 2020} \, ,$$ and $$H_1: \mbox{cases start date} < \mbox{March 1, 2020} \, ,$$ respectively. Only the estimated upper bound dates that were not discarded in the comparison with the community transmission data are considered in the test, providing a piece of strong evidence against the null hypothesis, with a p-value 0.0288.

This result is indirect evidence, coming from a prediction with a mathematical model, in favor of the thesis that the virus was already circulating in the city before March 2020, which adds to the direct (but weak) evidence that appears in the record of cases by first symptoms.

\subsection*{The appearance of other waves of contagion}

The exact periods when the other waves of contagion begin are so uncertain as that one of the first wave. Thus, periods that contain the other waves are determined by visual inspection as follows: Jun-Nov 2020 (second wave); Nov 2020-Mar 2021 (third wave); Mar-May 2021 (fourth wave); May-Jul 2021 (fifth wave); Jul-Dec 2021 (sixth wave). Logistic curves that fit training data, determined within each of these time windows, are fitted following a procedure similar to that adopted in the analysis of the first wave.

The results of these model calibrations can be seen in Table~\ref{tab2}, which shows for the 2nd to the 6th epidemic waves, and several peak date scenarios, information related to identified parameters, wave starting dates, fitting metric, and model selection criteria. Selecting the models as described in the first wave, we obtain the following models as the most representative: $\tau = 268$ (second wave); $\tau = 355$ (third wave); $\tau = 465$ (fourth wave); $\tau = 505$ (fifth wave); $\tau = 608$ (sixth wave).

\begin{figure*}
    \centering
    \includegraphics[scale = 0.3]{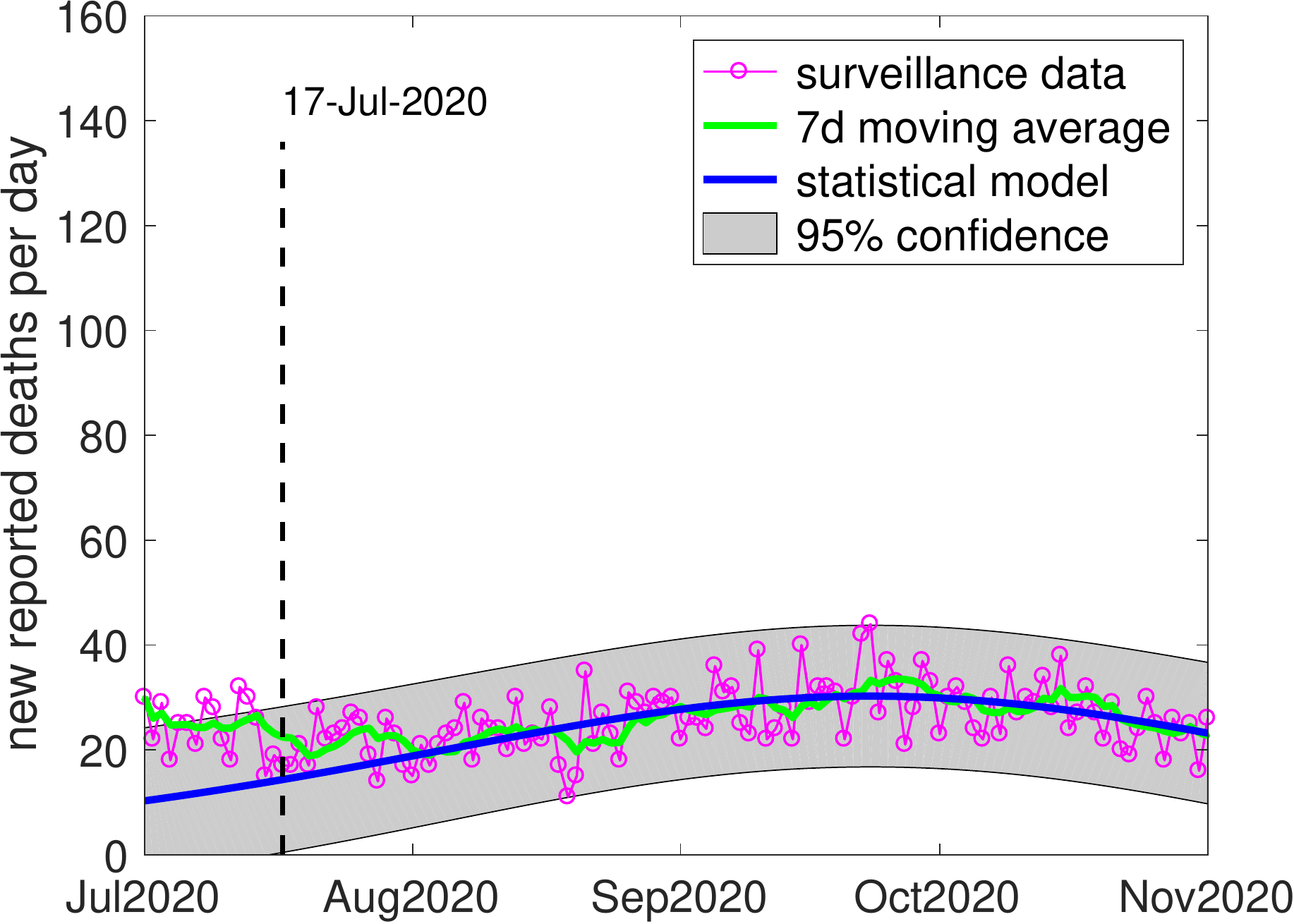}    
    \includegraphics[scale = 0.3]{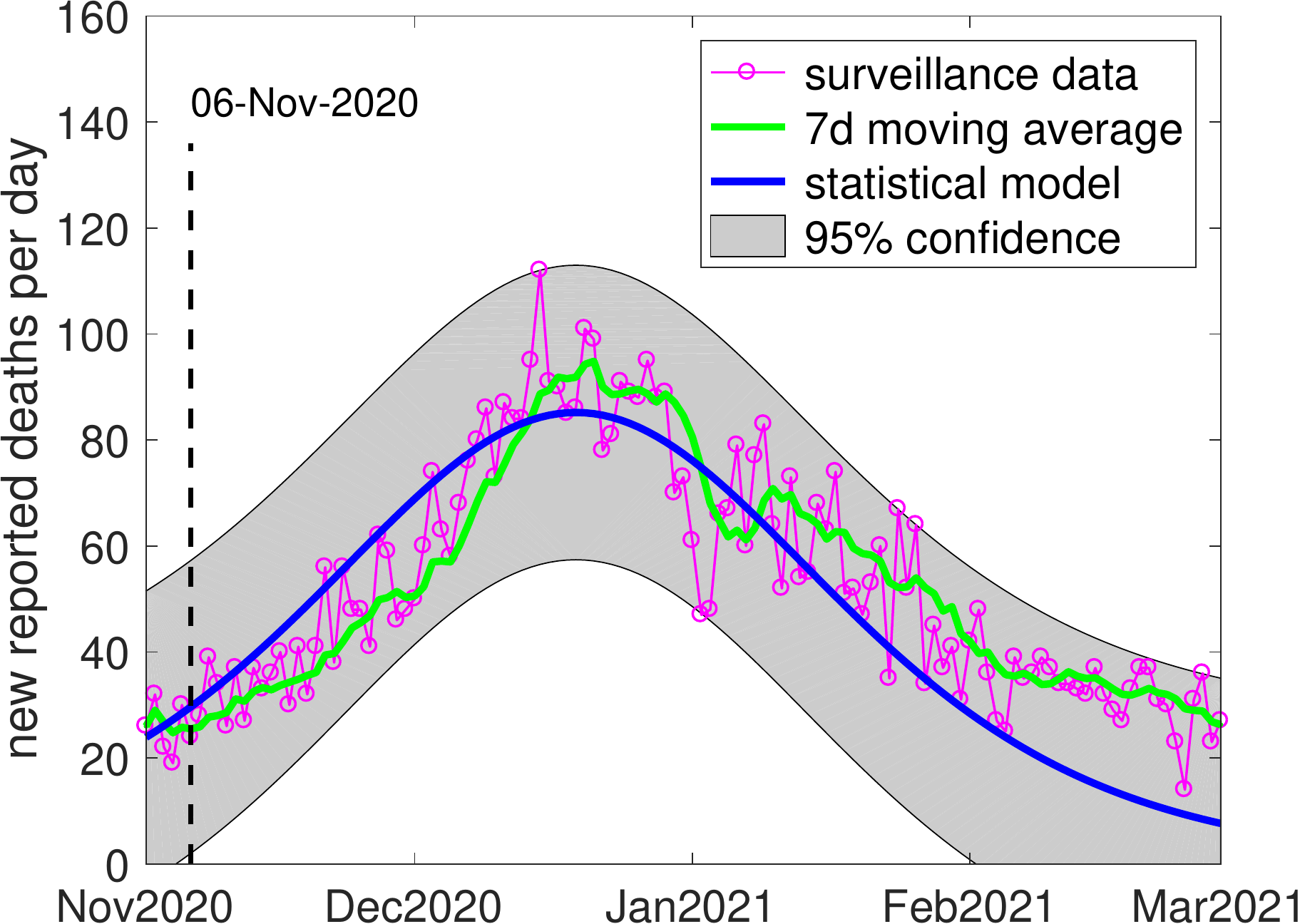}
    \vspace{5mm}
    
    \includegraphics[scale = 0.3]{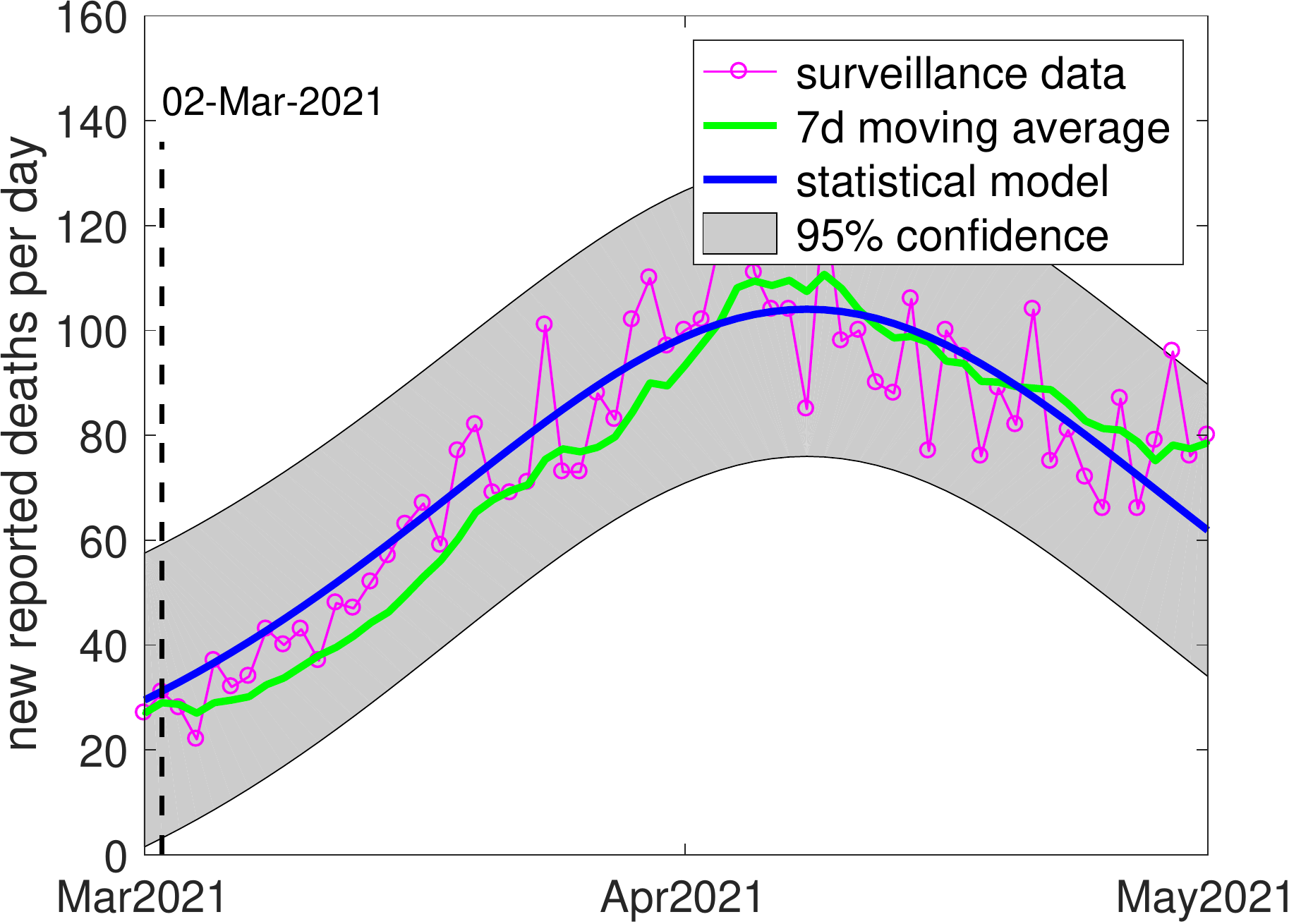}
    \includegraphics[scale = 0.3]{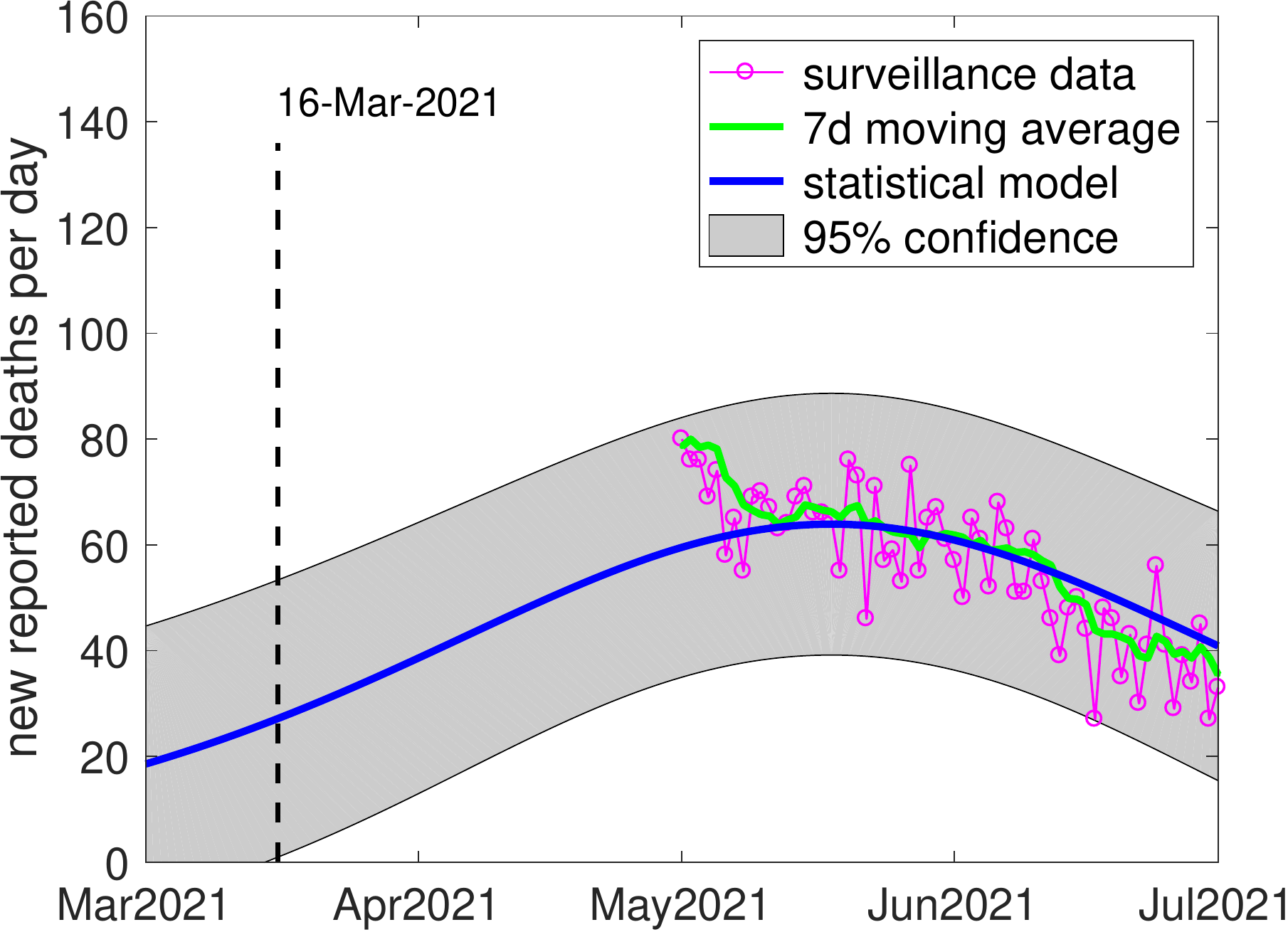}
    \includegraphics[scale = 0.3]{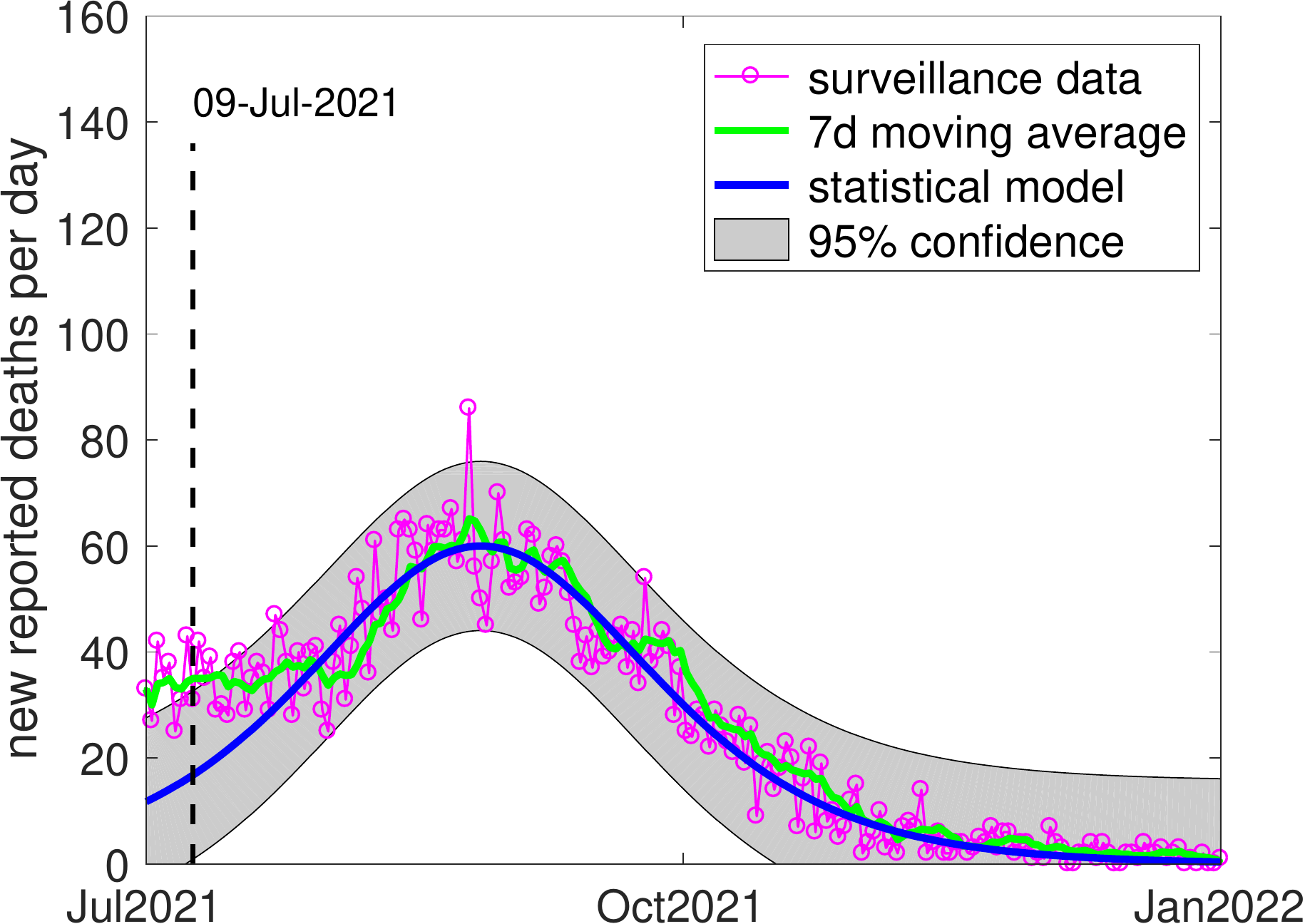}    
    \caption{Logistic curves fitted to the time series associated with COVID-19 incidence of deaths in Rio de Janeiro city, and the corresponding prediction bands, for several contagion waves. The dashed vertical lines indicate the upper bound estimation for the starting dates of the epidemic waves. Second wave (top) data: August 1st to November 1st, 2020; Third wave (middle top) data: November 1st, 2020 to February 1st, 2021; Fourth wave (middle middle) data: March 1st to May 1st, 2021; Fifth wave (middle bottom) data: May 1st to July 1st, 2021; and Sixth wave (bottom) data: August 1st to December 1st, 2021.}
    \label{fig:fig7}
\end{figure*}

The incidence curves that correspond to the best models of each wave can be seen in Fig.~\ref{fig:fig7}. In all these cases, the incidence curves show good adherence to the data, so that the respective models provide a reasonable description of the epidemic behavior underlying each analyzed period. Using these models (in the same way as was done in the first wave), we have as upper bounds for the start dates of the other epidemic waves: July 17, 2020 (second wave); November 6, 2020 (third wave); March 2, 2021 (fourth wave); March 16, 2021 (fifth wave); July 9, 2021 (sixth wave).

\begin{figure*}
    \centering
    \includegraphics[scale = 0.5]{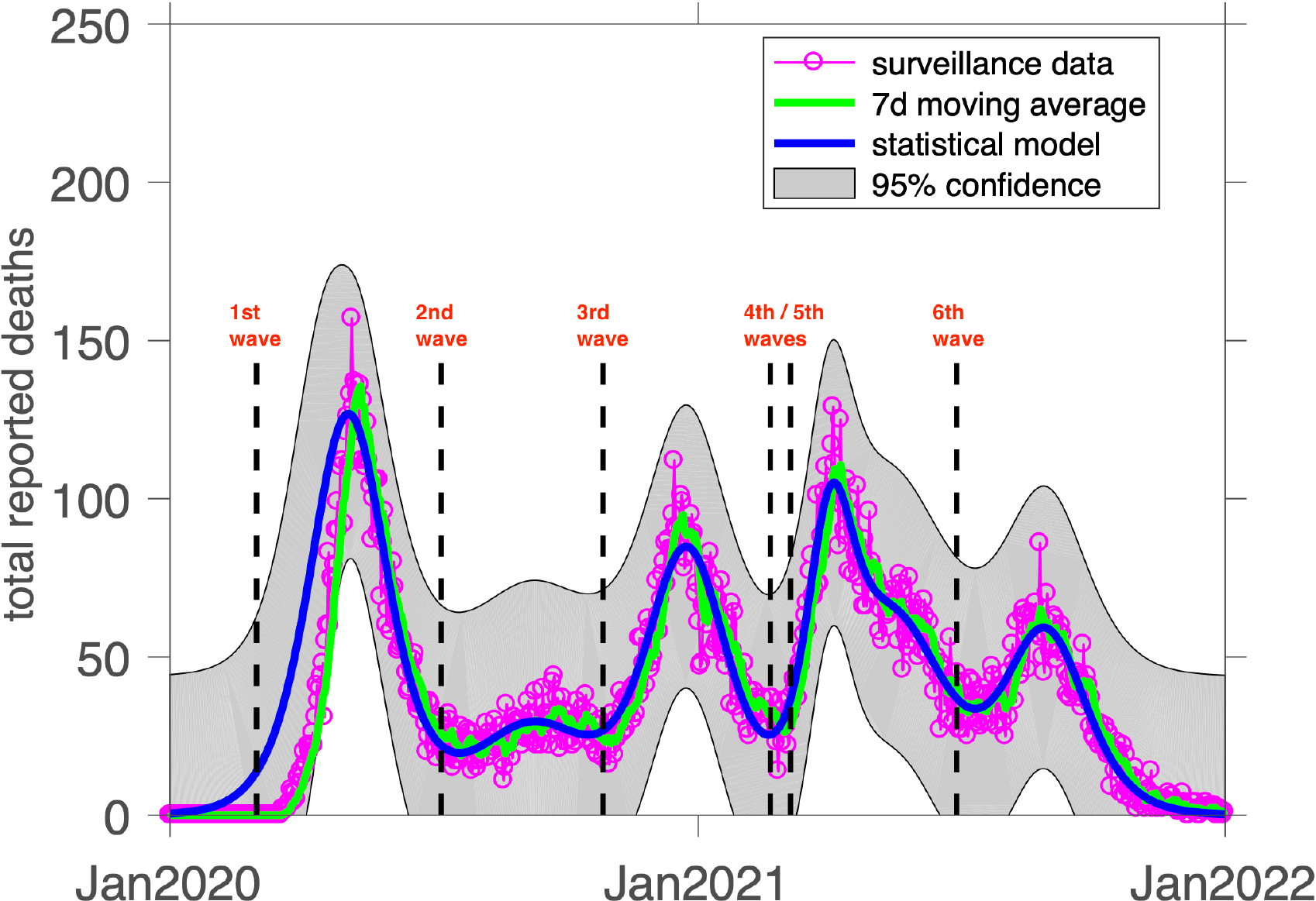}
    \caption{Logistic curve with 6-waves fitted to the time series associated with COVID-19 incidence of deaths in Rio de Janeiro city, and the corresponding prediction band. The dashed vertical lines indicate the upper bound estimations for the starting dates of the epidemic waves. Training data: April 1st, 2020 to December 31st, 2022.}
    \label{fig:fig8}
\end{figure*}

Using the parameters identified in each of the six waves above as the initial guess of a regression process that seeks to calibrate a model with 6 waves, Eq.(\ref{INt}), we obtain the incidence curve and prediction band that are shown in Fig.~\ref{fig:fig8}, which also shows the start dates that were estimated above. This figure clarifies a point that may have left doubts in the reader, about the temporal proximity between the fourth and fifth waves. The sum of the two can explain the asymmetry observed in the fall of the fourth wave, which is interpreted in the light of this model as being the superposition of two contagion waves that emerged temporally close. This is a plausible scenario, for example, if there are two key events in the city that are triggers for scaling up contagion events. Or if two regions of the city, geographically not close, experience an increase in contagion almost simultaneously.

\begin{table*}
\centering
\caption{{\bf Estimated parameters, the respective confidence intervals, fitting metrics, and information criteria measures for the statistical models used to represent the 2nd to 6th waves of deaths by COVID-19 in Rio de Janeiro (for several scenarios of peak day). 
}}
\label{tab2}
\vspace{2mm}
\begin{tabular}{c|c|c|c|c|c|cc|cc}
\toprule
 & $\tau$ & $K \times 10^3$ & $r \times 10^{-3} $ & deaths start date & cases start date & RMSE & R$^2$ & AIC & BIC\\
 & (day)	   & (people)               & (day$^{-1}$)             & upper bound      & upper bound      &         &            &        &      \\
\midrule
\multirow{4}{*}{2nd wave} 
& 264	   & 4.30 (3.63, 4.70) & 28 (23, 34) & Jul 16, 2020 & Jul 05, 2020 & 5.37 & 0.27 & 124.7 & 129.73\\
& 266	   & 4.36 (3.69, 5.03) & 28 (23, 33) & Jul 17, 2020 & Jul 06, 2020 & 5.34 & 0.27 &   88.2 &   93.30\\
& 268	   & 4.48 (3.78, 5.18) & 27 (22, 32) & Jul 17, 2020 & Jul 06, 2020 & 5.35 & 0.27 &     4.3 &     9.36\\
& 270	   & 4.67 (3.90, 5.43) & 26 (21, 31) & Jul 16, 2020 & Jul 05, 2020 & 5.40 & 0.26 &   22.8 &   27.84\\
\midrule
\multirow{4}{*}{3rd wave} 
& 349	   & 6.85 (6.22, 7.48) & 47 (41, 53) & Nov 06, 2020 & Oct 26, 2020 & 15.5 & 0.52 & 4.02 & 9.09\\
& 351	   & 6.66 (6.15, 7.17) & 50 (45, 55) & Nov 06, 2020 & Oct 26, 2020 & 13.7 & 0.62 & 4.00 & 9.07\\
& 353	   & 6.57 (6.14, 7.00) & 47 (47, 56) & Nov 06, 2020 & Oct 26, 2020 & 12.2 & 0.70 & 4.00 & 9.07\\
& 355	   & 6.54 (6.16, 6.92) & 48 (48, 56) & Nov 06, 2020 & Oct 26, 2020 & 11.0 & 0.76 & 4.02 & 9.11\\
\midrule
\multirow{4}{*}{4th wave} 
& 461	   & 6.01 (5.56, 6.47) & 70 (63, 77) & Mar 03, 2021 & Feb 20, 2021 & 13.2 & 0.75 & 4.00 & 8.25\\
& 463	   & 6.12 (5.73, 6.52) & 69 (63, 74) & Mar 02, 2021 & Feb 19, 2021 & 11.3 & 0.82 & 4.00 & 8.25\\
& 465	   & 6.36 (5.95, 6.77) & 65 (60, 71) & Mar 02, 2021 & Feb 19, 2021 & 10.9 & 0.83 & 4.00 & 8.25\\
& 467	   & 6.73 (5.25, 7.21) & 61 (55, 66) & Mar 02, 2021 & Feb 19, 2021 & 11.6 & 0.81 & 4.00 & 8.26\\
\midrule
\multirow{4}{*}{5th wave} 
& 504	   & 8.11 (6.78, 9.44) & 32 (26, 39) & Mar 14, 2021 & Mar 03, 2021 & 09.4 & 0.55 & 68.88 & 73.13\\
& 505	   & 8.11 (6.65, 9.57) & 32 (25, 38) & Mar 16, 2021 & Mar 05, 2021 & 09.7 & 0.52 & 11.52 & 15.78\\
& 506	   & 8.11 (6.51, 9.72) & 31 (24, 39) & Mar 19, 2021 & Mar 08, 2021 & 10.0 & 0.48 & 16.91 & 21.16\\
& 507	   & 8.11 (6.35, 8.87) & 31 (23, 39) & Mar 21, 2021 & Mar 10, 2021 & 10.3 & 0.45 & 37.77 & 42.02\\
\midrule
\multirow{4}{*}{6th wave} 
& 600	   & 4.86 (4.62, 5.10) & 49 (46, 52) & Jul 05, 2021 & Jun 24, 2021 & 6.3 & 0.92 & 4.27 & 9.89\\
& 602	   & 5.06 (4.81, 5.31) & 47 (44, 50) & Jul 02, 2021 & Jun 21, 2021 & 6.5 & 0.91 & 4.33 & 9.95\\
& 604	   & 4.86 (4.63, 5.10) & 49 (46, 52) & Jul 05, 2021 & Jun 24, 2021 & 6.3 & 0.92 & 4.27 & 9.89\\
& 608	   & 4.70 (4.48, 4.92) & 51 (48, 54) & Jul 09, 2021 & Jun 28, 2021 & 6.3 & 0.92 & 4.07 & 9.69\\
\bottomrule
\end{tabular}
\end{table*}

\subsection*{Correlation between starting dates and key events}

It is worth remembering at this point that the Carnival festivities of that year took place between the 21st and 26th of February (Friday to Wednesday) 2020. This temporal proximity between the plausible period for the beginning of community transmission (February 29 or before) and the Carnival is a shred of evidence in favor of the hypothesis widely circulated among epidemiologists\cite{Ashktorab2021} that the Carnival celebrations in the streets were a critical event for the spread of COVID-19 in Rio de Janeiro. Correlations with other key events, such as business reopenings, extended holidays, festive dates, etc., can be similarly investigated. Due to space limitations, we do not present other investigations of this type in this paper.

% --------------------------------------------------------------

% --------------------------------------------------------------
\section{Conclusions}
\label{sec:Conclusions}

The dynamics of COVID-19 may have multiple waves of contagion in the same geographic location as a side effect. Due to the complex and not fully understood nature of the disease transmission process, specifying the exact moment of emergence of new community contagion outbreaks is a hard task. This work presented a statistical approach capable of describing the multiple waves of contagion observed, as well as providing an interval estimation of the possible start date for each outbreak. 

This methodology was illustrated with the COVID-19 data of Rio de Janeiro city and determined possible starting dates for the multiple outbreaks during the years 2020 and 2021. The results obtained show that logistic models with one or more waves can be used to provide mathematical descriptions, with good adherence to the data, in epidemiological scenarios with complex transmission dynamics. In addition, they support that SARS-CoV-2 probably started to be disseminated locally in Rio de Janeiro as early as February 2020.
% --------------------------------------------------------------

% --------------------------------------------------------------
% rev p/ Americo
\section*{Supplementary Material}

The supplementary material is composed of the data and codes used to calibrate the statistical models, which are also are available in the repository \url{https://github.com/americocunhajr/COVID19Waves}

% --------------------------------------------------------------

% --------------------------------------------------------------
% rev p/ Americo
\begin{acknowledgments}

The authors thank Mrs. Gabrielle Pereira and \emph{\mbox{COVID-19:} Observatório Fluminense} initiative for preparing Figures~\ref{fig:fig1} and \ref{fig:fig2}. This work was supported by Funda\c c\~ao de Amparo \`a Pesquisa do Estado de S\~ao Paulo, FAPESP (process:  2015/50122-0); Funda\c c\~ao Carlos Chagas Fillho de Amparo \`a Pesquisa do Estado do Rio de Janeiro, FAPERJ (processes: 211.037/2019 and 201.294/2021); Coordenação de Aperfei\c coamento de Pessoal de N\'ivel Superior, CAPES (process:  88887.506931/2020-00), and Conselho Nacional de Desenvolvimento Cient\'ifico e Tecnol\'ogico , CNPq (process:  441016/2020-0).
\end{acknowledgments}
% --------------------------------------------------------------

% --------------------------------------------------------------
% rev p/ Americo
\section*{Data Availability Statement}
\label{sec:DataStatement}

The data that support the findings of this study  are available within the article and its supplementary material.

%Data openly available in a public repository that issues datasets with DOIs

%The data that support the findings of this study are openly available in [repository name] at http://doi.org/[doi], reference number [reference number].

% Data openly available in a public repository that does not issue DOIs

%The data that support the findings of this study are openly available in [repository name], reference number [reference number].
% --------------------------------------------------------------

% references
% --------------------------------------------------------------
\section*{References}

%\nocite{*}
%\bibliography{references}

%

% --------------------------------------------------------------

\end{document}